\documentclass[12pt]{article}

\usepackage[normalem]{ulem}

\usepackage{xcolor}
\usepackage[english]{babel}
\usepackage[T1]{fontenc}
\usepackage[utf8]{inputenc}
\usepackage{authblk}
\usepackage{mathtools}
\usepackage{slashed}
\usepackage{amsmath,amssymb}
\usepackage{amsfonts}
\usepackage{graphicx,color}
\usepackage{cite}
\usepackage{hyperref}
\hypersetup{
    colorlinks=true,
    linkcolor=blue,
    filecolor=magenta,      
    urlcolor=cyan,
}
\usepackage[capitalize]{cleveref}

\numberwithin{equation}{section} 

\definecolor{refcol}{rgb}{0.9,0.1,0.1}
\hypersetup{colorlinks=true,linkcolor=blue,citecolor=refcol,urlcolor=cyan,linktocpage}

\newcommand{\be}{\begin{equation}}
\newcommand{\ee}{\end{equation}}
\newcommand{\bea}{\begin{eqnarray}}
\newcommand{\eea}{\end{eqnarray}}

\def\Xint#1{\mathchoice
{\XXint\displaystyle\textstyle{#1}}%
{\XXint\textstyle\scriptstyle{#1}}%
{\XXint\scriptstyle\scriptscriptstyle{#1}}%
{\XXint\scriptscriptstyle\scriptscriptstyle{#1}}%
\!\int}
\def\XXint#1#2#3{{\setbox0=\hbox{$#1{#2#3}{\int}$ }
\vcenter{\hbox{$#2#3$ }}\kern-.6\wd0}}

\newcommand{\Qp}{\mathbb{Q}_p}
\newcommand{\Zp}{\mathbb{Z}_p}

\begin{document}
\begin{titlepage}
\thispagestyle{empty}

\title{
{\Huge\bf Matrix Model for Riemann Zeta}\\ 
{\Huge\bf via its Local Factors}
}

\vfill

\author{
    {\bf Arghya Chattopadhyay,}${}^{1}$\thanks{{\tt arghya.chattopadhyay@gmail.com}}\hspace{8pt}
    {\bf Parikshit Dutta}${}^{2}$\thanks{{\tt parikshitdutta@yahoo.co.in}},\\
    {\bf Suvankar Dutta}$^1$\thanks{{\tt suvankar.hri@gmail.com}}\hspace{8pt}
    {\bf and}\hspace{8pt}
    {\bf Debashis Ghoshal}${}^3$\thanks{{\tt d.ghoshal@gmail.com}} \\  
\hfill\\
\smallskip\hfill\\        
{\small
${}^1${\it Indian Institute of Science Education \&\ Research Bhopal,}\\
{\it Bhopal Bypass, Bhopal 462066, India}
\hfill\\
\smallskip\hfill\\        
${}^2${\it Asutosh College, 92 Shyama Prasad Mukherjee Road,}\\
{\it Kolkata 700026, India}
\hfill\\
\smallskip\hfill\\        
${}^3${\it School of Physical Sciences, Jawaharlal Nehru University,}\\
{\it New Delhi 110067, India}\\
and\\
{\it Albert Einstein Institute, Max Planck Institute for Gravitational Physics,}\\
{\it 14424 Potsdam, Germany
} }
     }

\vfill

\date{%
%
\begin{quote}
\centerline{{\bf Abstract}}
{\small
We propose the construction of an ensemble of unitary random matrices (UMM) for the Riemann zeta function. 
Our approach to this problem is `$p$-iecemeal', in the sense that we consider each factor in the Euler product representation 
of the zeta function to first construct a UMM for each prime $p$. We are able to use its phase space description to write 
the partition function as the trace of an operator that acts on a subspace of square-integrable functions on the $p$-adic 
field. This suggests a Berry-Keating type Hamiltonian. We combine the data from all primes to propose a Hamiltonian 
and a matrix model for the Riemann zeta function.}
\end{quote}
}

\vfill


\end{titlepage}
\thispagestyle{empty}\maketitle\vfill \eject

\tableofcontents

\section{Introduction}\label{sec:Introd}
The zeroes of the Riemann zeta function on the critical line do not appear to fall into any pattern, but rather seem to be 
randomly distributed (see section \ref{sec:Zeta} for details). The Euler product representation of the function (valid in a 
region of the complex plane) is believed 
to relate this to the apparent lack of pattern in the occurrence of the prime numbers. Montgomery \cite{montgomery} was 
the first to quantify this randomness through the two-point correlation function of the (non-trivial) zeroes of the zeta function. 
His serendipitous encounter with Dyson that helped in relating this to the behaviour of a random Gaussian ensemble of 
large unitary matrices is the stuff of a well known anecdote\cite{anecdote}. Since then extensive numerical work
\cite{Odlyzko} and analysis of higher correlators \cite{Hejhal:1994,sarnak1,Bogomolnyi:1995,Bogomolnyi:1996} (see also
\cite{Mehta2004}) have established this connection on a firm footing, and even extended to the closely related Dirichlet 
$L$-functions \cite{sarnak2}. All these are consistent with the suggestion of Hilbert and P\'{o}lya that these zeroes are in 
the spectrum of an unbounded, self-adjoint operator.  

Berry and Keating observed that the fluctuating part of the prime counting function, which is also related to the `height' of 
the non-trivial zeroes, resembles the Gutzwiller trace formula that gives, asymptotically, the fluctuating part of the energy 
distribution of a quantum Hamiltonian in terms of a sum over classically periodic orbits, along with corrections that depend 
on the deviations around them \cite{BK,BerryKeating2}. They, however, need to use the Euler product form of the zeta 
function in a domain where it is really not applicable. Be that as it may, the prime numbers correspond to the primitive 
periodic orbits and their powers to the repetitions. They propose that the non-trivial zeroes are the eigenvalues of a quantum 
dynamical system the classical Hamiltonian of which is $H=xp$. The authors caution that this cannot quite be, since the 
spectrum of this Hamiltonian, being related by a canonical transformation to the inverted harmonic oscillator, is not discrete. 
Moreover, the system is integrable, and hence cannot exhibit chaotic behaviour, which is an expected property of the 
Hamiltonian. Several possible resolutions were explored to overcome this difficulty. Among them are putting boundary or 
periodicity conditions on the coordinates and/or momenta. The dilatation symmetry, under which $x$ and $p$ scale so as 
to cancel each other, was also discussed. For further work on this model, with different boundary conditions, see Ref. 
\cite{Sierra:2007du}. Connes has also explored a Hamiltonian description for a much wider class of zeta functions 
\cite{ConnesEssay,ConnesTrForm}. This approach involve noncommutative geometry, trace formulas and $p$-adic 
number fields. It is, however, rather difficult to access for physicists. Ref.\cite{Schumayer:2011yp} has a comprehensive 
review and references of various physics related approaches to the Riemann zeta function.

Given an infinite set of distinguished points on a curve, e.g., the non-trivial zeroes of the Riemann zeta on the critical line, 
one can construct a random ensemble of unitary matrices (UMM) after conformally mapping the points on the 
unit circle. Using this approach, a unitary one-plaquette model was constructed for the symmetric zeta function (which is 
an entire function that has zeroes coinciding with the Riemann zeta) by two of us (PD and SD) in Ref.\cite{Dutta2016byx}. 
The parameters of the model are determined in terms of the Li (also known as the Keiper-Li) coefficients. Extending on 
earlier work \cite{Dutta:2007ws}, a phase space description of the model was developed. In this picture, an eigenvalue is 
the coordinate and the number of boxes in the Young diagram corresponding to a representation is the momentum. The 
phase space distribution function was computed and found to have a `broom-like' structure with spikes at the zeroes of 
the symmetric zeta function. While the `dynamical system'\footnote{The term is used in a general sense, since in a matrix 
model, we do not have a notion of time.}  is expected to be a many fermion system, we could not get any further insight 
from this phase space.  

In this paper, we shall follow a similar line of approach, however, we shall break it in parts. We shall take up each prime 
factor in the Euler representation of the Riemann zeta function (the so called local zeta function at a fixed prime) separately 
and construct an ensemble of unitary matrices (UMM), the eigenvalues of which correspond to its poles that lie on the 
imaginary axis. It turns out that the phase space density, or more accurately, the fluctuating part of it, can be expressed as 
the \emph{trace of an operator}. The operator in question is a generalised Vladimirov derivative that acts on a subspace of 
square integrable functions supported on the $p$-adic integers. This suggests that the natural notion of distance in the 
momentum space is ultrametric, defined in terms of the $p$-adic norm. The phase space picture also leads us to arrive 
at a Hamiltonian, which has the form $H=xp$. Interestingly its action is restricted to a subspace of the full domain --- this 
is reminiscent of a resolution to the shortcomings of this type of Hamiltonian in \cite{BK,BerryKeating2}. In this approach, 
even the classical phase space description is in terms of operators on a Hilbert space, as in the Koopman-von Neumann 
formalism \cite{wilczek,kvnthesis}. More specifically, the Hilbert space is a subspace spanned by the Kozyrev wavelets 
\cite{Kozyrev:2001} (an analogue of the generalised Haar wavelets) on the $p$-adic line. The restriction is to the subspace 
supported on the compact subset $\Zp$ of $\Qp$ with the translation parameter set to zero and scaling restricted so as to 
stay within the support. 

Finally, we attempt to combine the UMM of all primes (local data) to construct a random matrix model for the Riemann zeta 
function. (The part of the data that correspond to the trivial zeroes is related to matrix model for the gamma function, which 
could be thought of as a local zeta function at infinite place.) Not unexpectedly, this leads to a divergent expression for the 
coefficients of the model. We propose a \emph{renormalisation} procedure to deal with this divergence---this requires the 
introduction of one real parameter, which may be taken to be half in order to match with the critical line. This shift from the 
imaginary axis (relevant to the local zeta) to the critical line can be seen to arise from a similarity transformation of the 
Hamiltonian, although that too does not seem to fix the value of the renormalisation parameter. We end with a discussion on 
the nature of the phase space, and the Wigner function on it. 

The following observation is worth emphasising. Montgomery's correlation \cite{montgomery} of the zeroes imply that almost 
all the non-trivial zeroes of the zeta function are simple. Since the trivial zeroes are simple, all zeroes are  believed to be simple 
zeroes, i.e., of multiplicity one. This is actually true for at least 70\%\ of the non-trivial zeroes \cite{ConGhoGon,BuiHB}.  That 
two zeroes never come arbitrarily close is a property that resembles a special characteristics of fermions, which due to the 
Pauli exclusion, cannot occupy the same state. In an ensemble of matrices the eigenvalues behave like the position of fermions. 
Thus it is natural to seek a matrix model description for the zeroes of the zeta function.

We would like to stress that our objective is not to try to prove the Riemann hypothesis. Rather, our effort is to find a 
`dynamical system' through the UMM, by first constructing one for each prime factor, and then combining them. The motivation 
for this work is the Montgomery-Dyson observation relating the statistical distribution of the 
eigenvalues of an RMM and that of the non-trivial Riemann zeroes, augmented by recent advancement in our understanding
of the phase space underlying an RMM. This helped in the construction of a one-plaquette UMM starting with the non-trivial 
zeroes of the Riemann zeta function, which further gave a density function in the phase space \cite{Dutta2016byx}. In the present
work, we shall argue that the `$p$-iecemeal' approach mentioned above allows us to go a little farther towards this goal. It is our
hope that through this route one would get an idea about the `dynamical system' and the Hamiltonian, which could then be analysed
independently and perhaps more rigorously. It should be emphasised for this to work for the Riemann zeta function 
\cite{Dutta2016byx}, the zeroes must be on the line $\mathrm{Re }(s)=\frac{1}{2}$ (or, more precisely, on any curve that may then 
be mapped to the unit circle on the $z$-plane) as conjectured by Riemann. In contrast, the poles of the local zeta 
functions corresponding to the the prime factors are already on the line $\mathrm{Re }(s)=0$. We are aware that there 
are issues that will require better understanding, and we shall remark on these at appropriate places. 

The organisation in the following 
is as outlined in the table of contents. We begin in \cref{sec:Zeta} by recalling a few facts about the zeta function and 
$p$-adic analysis that we shall use subsequently. The UMM for the local zeta factors are constructed in \cref{sec:pUMM}, 
and their phase space picture developed in \cref{sec:PhaseSpace}. These results are combined in \cref{sec:Allp} in an 
attempt to build a UMM for the Riemann zeta function. We also discuss our proposal for a \emph{renormalisation} and 
attempt to understand this \emph{large phase space}. The resulting renormalised coefficients for the UMM have been 
compared In \cref{app:CompPS} with that for the symmetric zeta function constructed earlier in Ref.\cite{Dutta2016byx}. 

\section{Zeta functions and $p$-adic analysis}\label{sec:Zeta}
In this section, we shall recall a few facts about the Riemann zeta function as well as the $p$-adic number field and 
complex valued functions defined on it. We do not attempt to be comprehensive or even list the most important issues, 
but restrict to only those aspects that will be of importance for us in what follows.

\subsection{Riemann and local zeta functions}\label{ssec:RevZeta}
The infinite sum over the natural numbers
\begin{equation}
\zeta(s) = \sum_{n=1}^\infty \frac{1}{n^s}\label{EulerSum}
\end{equation}
for positive integer values of $s\ge 2$, was introduced by Euler, extended for $\text{Re}(s)> 1$ by Chebyshev and 
analytically continued to the complex $s$-plane as a meromorphic function by Riemann. This function has a simple pole 
at the origin, and vanishes for all negative even integers, which are known as its trivial zeroes. More interestingly, it 
has an infinite number of \emph{non-trivial} zeroes. Riemann also showed that $\zeta(s)$ satisfies the 
\emph{reflection identity}
\begin{equation}
\zeta(s) = 2^s \pi^{s-1} \sin\left(\frac{\pi s}{2}\right) \Gamma(1-s)\zeta(1-s)
\label{ReflId}
\end{equation}
from which one concludes that the non-trivial zeroes must all be in the \emph{critical strip} $0\le \text{Re}(s) \le 1$. 
However, he conjectured a much stronger result, namely that the non-trivial zeroes lie on the \emph{critical line} 
$\text{Re}(s)=\frac{1}{2}$. This is the celebrated \emph{Riemann hypothesis} that is among the most famous unsolved 
problems in mathematics. Associated to it is the \emph{symmetric zeta-function} 
\begin{equation}
\xi(s) = \frac{1}{2}\pi^{-s/2}s(s-1)\Gamma\left(\frac{s}{2}\right)\zeta(s) \label{SymmZ}
\end{equation}
which is an entire function that satisfies $\xi(s)=\xi(1-s)$, and has zeroes only at the non-trivial zeroes of $\zeta(s)$, 
at $s= \gamma_m =\frac{1}{2} + i t_m$, where $t_{m}$ could possibly be complex.

Li proved that the Riemann hypothesis is equivalent to the non-negativity of the following sequence of real numbers 
\cite{Li} (see also \cite{Keiper,Bombieri})
\begin{equation}
\label{LiNumbers}
\lambda_{n} = \frac{1}{(n-1)!}\,\frac{d^{n}}{ds^{n}}\Big[ s^{n-1}\ln\xi(s)\Big]_{s=1}
\end{equation}
called the \emph{Li coefficients} (sometimes the Keiper-Li coefficients). Using \cref{EulerProd}, these can be expressed as
\begin{equation}
\label{AltLiNumbers}
\lambda_{n} = \sum_{m}\left(1- \left(1-\frac{1}{t_m}\right)^{n}\right)
\end{equation}
These numbers appear in the parameters that define the matrix model associated with the zeta function \cite{Dutta2016byx}.

Euler also expressed the sum \cref{EulerSum} as an infinite product 
\begin{equation}
\zeta(s) = \prod_{p\,\in\,{\mathrm{primes}}} \frac{1}{\left(1 - p^{-s}\right)}\label{EulerProd}
\end{equation}
over the \emph{prime} numbers. It is natural to expect, therefore, that the distribution of the zeroes of the zeta-function, 
i.e., the locations of the $t_m$, to be related to the distribution of the primes.

The factors in \cref{EulerProd} are called the \emph{local zeta-function at the prime} $p$
\begin{equation}
\zeta_p(s) = \frac{1}{\left(1 - p^{-s}\right)}\label{pZeta}
\end{equation}
The local zeta function does not have any zero, rather it has equally spaced simple poles at $s = 
\displaystyle{\frac{2\pi i}{\ln p} n}$ for $n\in\mathbb{Z}$ on the vertical line $\text{Re}(s) = 0$. 

The \emph{prime counting function} $J(x) = \!\!\displaystyle{\sum_{p\,\in\,{\mathrm{primes}}} \sum_{n\in{\mathbb{Z}^+}}} \,
\frac{1}{n}\,\Theta(x - p^n)$ (where $\Theta(x)$ is the Heavi\-side step function) is a monotonically increasing function that 
jumps by $\frac{1}{n}$ at every power of prime numbers $p^n$\cite{hmedwards} (see also \cite{Zagier1977,Rudnick1996}
for more accessible introductions for physicists). It can be written as 
\begin{equation}
J(x) = \mathrm{Li}(x) - \sum_{m} \mathrm{Li}(x^{t_m}) -  \ln 2 + \int_{x}^{\infty} \frac{dt}{t(t^{2}-1)},
\qquad x>1 \label{PrimeCountJ}
\end{equation}
where 
$\displaystyle{\mathrm{Li}(x) =\int_0^{x}\frac{dt}{\ln t}}$
is the logarithmic integral function. Riemann showed that $\ln\zeta(s)$ and $J(x)$ are related by Mellin transformation
as follows.
\begin{align}
\label{RelnZetaJ}
\begin{split}
\ln \zeta(s) &= s \int_0^\infty \frac{dx}{x^{s+1}} J(x), \qquad \mathrm{Re}(s)>1\\
J(x) &= \frac{1}{2\pi i } \int_{a-i \infty}^{a+\infty} \frac{ds}{s}\, x^s\, \ln \zeta(s), \qquad a>1
\end{split}
\end{align}
The function $J(x)=\overline{\!J}(x) + \widetilde{J}(x)$, has an average part $\overline{\!J}(x)$ and a fluctuating part 
$\widetilde{J}(x)$, the latter may be expressed in terms of zeroes of the zeta function\cite{hmedwards} as
\begin{equation} 
d\widetilde{J}(x) = - \frac{dx}{\ln x}\, \sum_{m} x^{t_m-1}
\end{equation}
A related function that we shall find use for later is the summatory von Mangoldt function or the Chebyshev counting 
function\cite{hmedwards}
\begin{align}\label{sumVM}
\begin{split}
\psi(x) &= \displaystyle{\sum_{p\,\in\,{\mathrm{primes}}}\sum_{n\in {\mathbb{Z}^+}}} \,\ln p\;\Theta(x - p^n) \\
\frac{d\psi(x)}{dx} &= \sum_{p}\ln p\,\frac{d \jmath_{p}(x)}{dx}\; =\; \left(1 - \sum_{\gamma_m} x^{\gamma_m - 1} 
- \sum_{n=1}^{\infty} x^{-2n-1}\right) \quad\quad (x>1) 
\end{split}
\end{align}
In the above we have an defined $\jmath_p(x) = \displaystyle{\sum_{n\in{\mathbb{N}}}} \,\Theta(x - p^n)$, a 
\emph{local counting function} at a prime $p$,  for which
\begin{equation}
\int_0^\infty x^{-s} \frac{d\jmath_p(x)}{dx} dx = \sum_{n\in{\mathbb{N}}} p^{-ns} = p^{-s}\zeta_p(s) \label{locJZeta}
\end{equation}
Notice that $\jmath_p(x)$ jumps by 1 at the powers $p^n$ of the fixed prime $p$. The following identity proves very 
useful.
\begin{equation}
\sum_{n\in\mathbb{N}} \delta(q - n\ln p) = \frac{1}{\ln p} \sum_{m\in\mathbb{Z}} \exp\left(\frac{2\pi i q m}{\ln p}\right),
\quad \text{for }\, q > 0  \label{IdentityInfSum}
\end{equation}
In deriving this, we change variables as $t = 1-s$ and $x = e^q$ in the inverse Mellin transform of \cref{locJZeta} 
and pick up the poles at integer multiples of $2\pi i/\ln p$ in the $t$-plane. On the other hand, if we consider the sum 
$\displaystyle{\sum_{n=0}^\infty\delta(x - p^{-n})}$ and follow the same route, we get 
\begin{equation*}
\sum_{n=0}^\infty \delta(q + n\ln p) = \frac{1}{\ln p} \sum_{m\in\mathbb{Z}} \exp\left(\frac{2\pi i q m}{\ln p}\right),
\quad\text{for }\, q \le 0 
\end{equation*}
which, when combined with \cref{IdentityInfSum}, yields the Fourier transform of the Dirac comb 
\begin{equation}
\sum_{n\in\mathbb{Z}} \delta(q - n\ln p) =\frac{1}{\ln p} \sum_{m\in\mathbb{Z}} \exp\left(\frac{2\pi i q m}{\ln p}\right),
\quad\text{for }\, q \in\mathbb{R}  \label{ShaFT}
\end{equation}
Notice that although the supports of the $\delta$-function accumulate near the origin in the $x$-plane, they are well
separated in the $q$-plane.

\subsection{A few results from $p$-adic analysis}\label{ssec:RevPAdic}
The motivation for the local zeta function at prime $p$ comes from the $p$-adic field $\Qp$ and analysis on it. We shall 
need to use some results from there. Let us recapitulate a few facts about the field $\Qp$ and some aspects of $p$-adic 
analysis. More details are available at many places, e.g. Refs.\cite{koblitz1996p,VVZ1994p,robert2013}. 

First, let us fix a prime $p$. The $p$-adic norm of a rational number $m/n$ is $\left|\displaystyle{\frac{m}{n}}\right|_p = 
p^{\text{ord}_p(n) - \text{ord}_p(m)}$, where ord$_p$ is the highest power of $p$ that divides its argument. Starting with 
the rationals, the field $\Qp$ is the Cauchy completion obtained by including the limits of all converging sequences (upto 
an appropriately defined notion of equivalence) with respect to the $p$-adic norm. This norm has the ultrametric property 
leading to a stronger than usual triangle inequality $\left| \xi-\xi' \right|_p \leq \text{max}\left( |\xi|_p, |\xi'|_p\right)$. Since the 
real numbers are obtained by the same procedure, but using the absolute value norm, $\Qp$'s are similar to $\mathbb{R}$. 
Indeed, upto equivalence, the absolute value norm and the $p$-adic norms are the only possible norms on the set of 
rationals. 

An element $\xi\in\Qp$ admits a Laurent series expansion in $p$: 
\begin{equation}
\xi = p^{N}\left(\xi_0+\xi_1p+\xi_2p^2+\cdots\right) = p^{N}\sum_{n=0}^\infty\xi_n p^n,\qquad |\xi|_p = p^{-N} 
\label{plaurent}
\end{equation}
where $N\in\mathbb{Z}$ and $\xi_n\in\{0,1,\cdots,p-1\}$, but\footnote{There is, however, nothing special about this 
choice---one may work with other representative elements.} $\xi_0\ne 0$. The series above is convergent in the $p$-adic 
norm. The subset $\Zp = \left\{\xi\in\Qp\; :\; |\xi|_p\le 1\right\}$, i.e., elements with norm less than equal to 1, is the compact 
subring of $p$-adic integers. 

One can define functions on $\Qp$, which could be valued in any field. In particular, we shall be interested in complex 
valued functions. For example, $\chi:\Qp\to\mathbb{C}$, defined as
\begin{equation}\label{pexponential}
\chi^{(p)}(\xi) = \exp\left(2\pi i\xi\right)
\end{equation}
has the property $\chi^{(p)}(\xi+\xi')=\chi^{(p)}(\xi)\chi^{(p)}(\xi')$, and is called an {\it additive character} of $\Qp$. The 
totally disconnected nature of the $p$-adic space means that the (complex valued) continuous functions on $\Qp$ are 
locally constant. One such function is the \emph{indicator function} 
\begin{equation}
\mathfrak{G}^{(p)}(\xi-\xi_0) = \left\{ 
\begin{array}{ll}
1 & \text{for } \left|\xi-\xi_0\right|_p \le 1\\
0 & \text{otherwise}
\end{array}
\right. \label{indicator}
\end{equation}
and similarly for other open sets. In fact, the indicator functions on open sets are the $p$-adic analogue of the
Gaussian function on the real line, in the sense that these retain their form after a Fourier transformation. 

There is a translationally invariant Haar measure\footnote{In spite of the counter-intuitive topology of $\Qp$, its field 
structure allows for a translation (as well as a scale invariant) Haar measure. For details of integration of {\em complex 
valued} functions and distributions on $\Qp$, as well as their relation to Lebesgue integrals, see \cite{VVZ1994p}.} 
$d\xi$ on $\Qp$. The discrete valuation of $p$-adic numbers reduces the problem of integration to evaluation of sums. 
The measure is normalised by
\[
\displaystyle{\int_{\Zp}} d\xi=1
\]
The following integral, which is easy to evaluate,
\begin{equation}
\int_{p\Zp} \left| \xi\right|_p^{s-1} d\xi = \frac{p-1}{p}\,\frac{p^{-s}}{\left(1-p^{-s}\right)},\qquad 
p\mathbb{Z}_p = \left\{\xi\in\mathbb{Q}_p :\: \left| \xi\right|_p < 1 \right\} 
\label{eq:pInt}
\end{equation}
will be important for us in what follows. 

On the other hand, the usual notion of derivative does not work because of the totally disconnected topology 
of $\Qp$. The Vladimirov derivative \cite{VVZ1994p}, therefore, is defined as an integral kernel
\begin{equation}\label{VladD}
D^\alpha f(\xi) = \frac{1-p^\alpha}{1-p^{-\alpha-1}}\, \int d\xi'\, \frac{f(\xi')-f(\xi)}{|\xi'-\xi|_p^{\alpha+1}}
\end{equation}
which is in fact well-defined for any $\alpha\in\mathbb{C}$ \cite{Albeverio:2011}. Furthermore, 
$D^{\alpha_1} D^{\alpha_2} = D^{\alpha_2} D^{\alpha_1} = D^{\alpha_1+\alpha_2}$.

The additive character \cref{pexponential} and the indicator function \cref{indicator} are the ingredients for a set of 
functions defined by Kozyrev in \cite{Kozyrev:2001}
\begin{equation}
\psi_{n,m,j}^{(p)} (\xi) = p^{-\frac{n}{2}} \chi^{(p)} \left( j p^{n-1} \xi \right) \mathfrak{G}^{(p)} \left(p^n \xi - m \right), 
\quad \xi \in \Qp 
\label{pWavelet} 
\end{equation}
for $n\in\mathbb{Z}$, $m \in \Qp/\Zp$ and $j \in \{1,2,3,\cdots,p-1\}$. They provide an orthonormal basis for 
$L^2(\Qp)$ 
\begin{equation}
\int_{\Qp} \psi_{n,m,j}^{(p)}(\xi) \psi_{n', m',j'}^{(p)}(\xi)\, d\xi = \delta_{n,n'}\delta_{m,m'}\delta_{j,j'}
\label{orthognality}
\end{equation}
Moreover, they satisfy
\begin{eqnarray}
\int_{\Qp} \psi_{n,m,j}^{(p)}(\xi) \, d\xi = 0
\end{eqnarray}
These properties are exactly analogous to the generalised Haar wavelets on $\mathbb{R}$. In addition, the Kozyrev 
functions are eigenfunctions of the Vladimirov derivative
\begin{equation}
D^\alpha \psi_{n,m,j}^{(p)} (\xi) = p^{\alpha(1-n)} \psi_{n,m,j}^{(p)} (\xi) \label{VDonKozy}
\end{equation}
with eigenvalue $p^{\alpha(1-n)}$. As shown in Ref.~\cite{DuGhLa}, this enhances the scaling symmetry of the 
$p$-adic wavelets to a larger SL(2,$\mathbb{R}$) group .

It is possible to restrict the wavelets to a compact subset\footnote{The support of the Kozyrev wavelets and their orbits 
under scaling and translations have been explained in detail in \cite{DuGhLa}.}, say $\Zp$, just as the Haar wavelets on 
the real line may be restricted to the finite interval $\left[0,1\right]$. This would mean restricting scaling of the mother 
wavelet by the parameter $n$ only in negative integers (only contractions), and the translation parameter $m$ can take 
a finite number of values that satisfy $|m|_p = p^{-n}$ (the parameters of translation are in $\Qp/\Zp$). Restricted thus, 
the wavelets span a subspace $L^2(\Zp) \subset L^2(\Qp)$.

Before we close this section, let us get back to the zeta function. The Euler product formula shows that the local 
zeta functions \cref{pZeta} are naturally associated with the prime numbers, and by extension to $\Qp$. On the
other hand, $\mathbb{R}$ is the only other completion of the rationals. The symmetric zeta function \cref{SymmZ}
motivates the following definition
\begin{equation}
\zeta_{\mathbb{A}}(s) = \pi^{-\frac{s}{2}}\Gamma\left(\frac{s}{2}\right)\,\prod_p\zeta_p(s)
\equiv \zeta_{\mathbb{R}}(s)\prod_p\zeta_p(s)\label{adelicZ}
\end{equation}
called the \emph{adelic zeta function} that  involves the zeta functions for $\Qp$ for all primes and for $\mathbb{R}$. 
It satisfies $\zeta_{\mathbb{A}}(s)=\zeta_{\mathbb{A}} (1-s)$. It is interesting to note that
\begin{equation}
\zeta_{\mathbb{R}}(s) = \pi^{-\frac{s}{2}}\Gamma\left(\frac{s}{2}\right) = \int_{\mathbb{R}} dx\,\left| x \right|^{s-1} 
e^{-\pi x^2}\label{zetaRMellin}
\end{equation}
that is, the LHS is the Mellin transform of the Gaussian function. On the other hand, the local zeta function
\begin{equation}
\zeta_p (s) = 
\frac{p}{p-1} \int_{\Qp} dx\,\left| x \right|^{s-1} \mathfrak{G}_p(x) \label{zetaQpMellin}   
\end{equation}
is also a Mellin transform of the $p$-adic analogue of the Gaussian \cite{Brekke}. Therefore, in a sense 
$\pi^{-\frac{s}{2}}\Gamma\left(\frac{s}{2}\right) = \zeta_{\mathbb{R}}(s) \equiv \zeta_\infty(s)$ is the 
\emph{local factor} related to $\mathbb{R}$. It is also called the zeta function at \emph{infinite place}.   

The \emph{adelic} ring $\mathbb{A} = \mathbb{R}\times\prod_p\Qp$ consists of elements $x = \left(x_\infty, x_2, 
x_3, x_5,\cdots\right)$, where $x_\infty\in\mathbb{R}$, $x_p\in\Qp$, and all but a finite number of $x_p$ are $p$-adic 
integers ($|x|_p \le 1$). Elements can be added and multiplied component by component. An example is $n = (n, n, n, 
\cdots), n\in\mathbb{Z}$, for which the adelic relation $|n| \displaystyle{\prod_{p\in{\mathrm{primes}}}} \left| n \right|_p = 1$, 
relating the product of all $p$-adic norms together with the absolute value norm of an integer, holds. We shall not need 
the ad\`{e}le, but would like to point out that that the term in \cref{adelicZ} is motivated by it. 

\section{Unitary matrix model for local zeta}\label{sec:pUMM}
A unitary matrix model for the Riemann zeta function was constructed in \cite{Dutta2016byx}. To be precise, the matrix 
model constructed there was for the logarithm of the symmetric zeta function. In this section, we shall construct a matrix 
model corresponding to the local zeta functions. We shall largely follow the procedure developed there, however, we shall 
work directly with the function $\zeta_p(s)$ and not $\ln\zeta_p(s)$. More precisely, our objective will be to construct an 
ensemble of unitary matrices, the eigenvalue distribution of which coincide with the distribution of the poles of a local zeta 
function. Since the eigenvalues of unitary matrices lie on the unit circle, we first map the poles of \cref{pZeta} on the 
imaginary $s$-axis to the unit circle on the $z$-plane by the conformal map
\begin{equation}
z = \frac{s-1}{s+1},\qquad s = \frac{1+z}{1-z} = i\cot\frac{\theta}{2} \label{ConfMap}
\end{equation}
The partition function of a generic one plaquette unitary matrix model is
\begin{equation}
Z = \int {\cal{D}}U\, \exp\left(N\sum_{n=1}^{\infty} \frac{\beta_{n}}{n} \left(\text{Tr } U^{n} + \text{Tr } U^{\dagger n}\right)
\right)\label{PFPlaq}
\end{equation}
where the real numbers $\beta_n$ are the parameters of the theory. In the following, we shall find
the set of parameters $\{\beta_n\}$ corresponding to the poles of $\zeta_p(s)$.

\subsection{Eigenvalue distribution}\label{ssec:EValDis}
An eigenvalue analysis of this model has been done in \cite{Jurkiewicz:1982iz,Mandal:1989ry}. Consider an ensemble 
of $N\times N$ matrices, the partition function can be rewritten by change of variables to the eigenvalues
\begin{equation}\label{eq:pfeige}
Z = \int \prod_{i=1}^N d\theta_i\, \prod_{i < j} 4 \sin^2 \left(\frac{\theta_i - \theta_j}{2}\right)\,
\exp\left(  N \sum_{n=1}^{\infty} \sum_{i=1}^N \frac{2\beta_n}{n} \cos n\theta_i \right)
\end{equation}
The additional factor, known as the Vandermonde determinant, is due to the Jacobian of the transformation. In the limit 
$N\to\infty$, let us define the continuous variables $x$ and $\theta(x)$ as 
\begin{equation}
x = \frac{i}{N} \in [0,1],\qquad \theta_i \rightarrow \theta(x) \nonumber
\end{equation}
to arrive at the partition function
\begin{align}
\label{eq:pfeige2}
\begin{split}
Z &= \int \mathcal{D}\theta\, \exp\left({-N^2 S_{\mathrm{eff}}[\theta]}\right)\\
S_{\mathrm{eff}}[\theta] &= - \sum_{n=1}^{\infty} \frac{2\beta_n}{n} \int_0^1 dx \cos n\theta(x) - 
\frac{1}{2} \int_0^1\!dx\;\Xint-_0^1 dy \ln \left|4 \sin^2 \left(\frac{\theta(x)-\theta(y)}{2}\right)\right|
\end{split}
\end{align}
where, the Jacobian has been incorporated in the effective action $S_{\mathrm{eff}}[\theta]$. 

In the large $N$ limit, the dominant contribution comes from extrema of $S_{\mathrm{eff}}[\theta]$ which is determined by 
the saddle point equation
\begin{equation}\label{eq:eveqnplaq} 
\Xint-_{-\pi}^{\pi} d\theta'\,\rho(\theta') \cot\left({\frac{\theta - \theta'}{2}}\right) = V'(\theta)
\end{equation}
where,
\begin{equation}\label{eq:V'def}
\rho(\theta) = \frac{\partial x}{\partial\theta} \quad 
\text{and} \quad V'(\theta) = 2\sum_{n=1}^{\infty} \beta_n \sin n\theta
\end{equation}
Thus for a given set of the parameters $\beta_n$, one has to solve saddle point equation \cref{eq:eveqnplaq} to find the 
density of eigenvalues $\rho(\theta)$.

\subsection{The resolvent}\label{ssec:Rzprop}
It is difficult, in general, to solve the integral equation \cref{eq:eveqnplaq} to determine the distribution of the eigenvalues. 
It is easier to work in terms of the \emph{resolvent}
\begin{equation} \label{eq:Rz}
R(z) = \frac{1}{N}\,\left\langle \text{Tr}\left(\frac{1}{1-zU}\right)\right\rangle
\end{equation}
that is analytic both inside, as well as outside, the unit circle. Expanding in Taylor series around $z=0$ and $\infty$ in the 
two regions 
\begin{equation}\label{rzexpan}
R(z) = \left\{ \begin{array}{ll}
1 + \frac{1}{N} \left( z \left\langle\text{Tr} U\right\rangle + z^2  \left\langle\text{Tr} U^2\right\rangle 
 + \cdots\right) & \: \text{for }\left|z\right|<1\\
 {} & {}\\
- \frac{1}{N} \left( \frac{1}{z} \left\langle\text{Tr} U\right\rangle + \frac{1}{z^2}  \left\langle\text{Tr} 
U^2\right\rangle  + \cdots\right) & \: \text{for }\left|z\right|>1 
\end{array} \right.
\end{equation}
Since $\frac{1}{N}\left\langle\text{Tr} U^k\right\rangle$ lies between $-1$ and $+1$ for all $k$, the Taylor series is convergent. 
The resolvent should be such that 
\begin{align}
R(0)=1,\: &\quad R(z\to\infty)=0 \nonumber\\  
\text{and }\, R(z) &+ R\left(\frac{1}{z}\right) = 1 \label{eq:Rzprop1}  
\end{align}
All these are satisfied by \cref{rzexpan}. It should be emphasised that \cref{eq:Rzprop1} is sufficient to show that 
the resolvent $R(z)$ solves a Riemann-Hilbert problem. The imaginary part of $R(z)$ is equal to the derivative of the potential of
the matrix model potential. The connection between the solutions of Riemann-Hilbert problem and resolvents coming from different 
matrix models allows one to use this mechanism in reverse. Thus, starting with a function which is complex analytic everywhere
(except on a specified curve), satisfies the conditions in \cref{eq:Rzprop1} and  solves a Riemann-Hilbert problem, one may view it 
as the resolvent of some matrix model. In this paper we will follow this reverse mechanism to construct different unitary matrix models, 
corresponding to different prime numbers, starting from appropriate complex functions tailored to our specific needs. Further, in the 
large $N$ limit the resolvent satisfies an algebraic (quadratic) equation, rather than an integral equation, that is far easier to solve.

In order to obtain this equation one uses the fact that the Haar measure is invariant under a more general transformation 
$U \rightarrow U e^{t A}$, where $A$ is an anti-symmetric matrix and $t$ is a \emph{real} parameter. The partition function 
\cref{eq:pfeige2}, being an integral over all unitary matrices, is also invariant under this transformation. This yields the identity
\begin{equation}
\frac{d}{dt}\int\!\! {\cal{D}}U\, \frac{1}{N}\text{Tr}(1-zUe^{t A})^{-1} \exp\left(\! N\sum_{n=1}^{\infty} \frac{\beta_{n}}{n}
\left( \text{Tr}(Ue^{tA})^{n} + \text{Tr} (U^{\dagger}e^{-tA})^{n}\right)\right)\Bigg|_{t=0} = 0
\end{equation}
which is known as the Dyson-Schwinger equation.

Following the steps in \cite{Friedan:1980tu,Dutta2016byx}, a solution to this equation may be written in the form
\begin{equation}\label{SolnRZed}
R(z) = \frac{1}{2} \left( 1+\sum_{n=1}^{\infty}\beta_{n} \left( z^{n}-\frac{1}{z^{n}}\right) \pm \sqrt{F(z)}\right)
\end{equation}
where the choice of the sign $+$ or $-$ is for $|z|$ less, respectively greater, than 1, and the function $F(z)$ is 
\begin{align}\label{eq:EffZed}
\begin{split}
F(z) &= \left( 1 + \sum_{n=1}^{\infty} \beta_{n} \left( z^n + \frac{1}{z^{n}}\right)\right)^{2} -
4 \left( \sum_{n=1}^{\infty} \beta_{n}z^{n}\right)\left( \sum_{m=1}^{\infty}\frac{\beta_{m}}{z^{m}}\right) \\
&\qquad + 4\beta_{1} R'(0) + 4\beta_{2} \left( \frac{R'(0)}{z} + \frac{1}{2} {R''(0)} + R'(0)z\right)\\
&\qquad + 4\beta_{3} \left( \frac{R'(0)}{z^{2}} + \frac{R''(0)}{2z} + \frac{1}{3!} R'''(0) + \frac{1}{2} R''(0) z +
R'(0) z^{2} \right) + \cdots
\end{split}
\end{align}
The analyticity properties of $R(z)$ inside the unit circle depend on the form of $F(z)$, which determines the phase structure 
of the model (see \cite{Dutta:2015noa} for more details). In the `no-gapped phase' (in which the eigenvalues are distributed 
over the entire unit circle), only the first terms in the parentheses survive, therefore, $F(z)$ is a perfect square and 
\begin{equation}
R(z) = 1 + \displaystyle{\sum_{n=1}^\infty}\beta_n z^n\qquad\text{(in the no-gapped phase)}
\label{NoGapRes} 
\end{equation}
from which one can find $\beta_n$ if the resolvent is known.

It is not difficult to surmise a resolvent function that satisfies all the requirements outlined above, in particular \cref{eq:Rzprop1}, 
and carries information about the poles of the local zeta function. We propose to work with
\begin{equation}
R^{(p)}(z) = 
\begin{cases}
\begin{array}{l}
1 - \displaystyle{\frac{z}{2(1-z)^{2}}\, \frac{p^{\frac{1+z}{2(1-z)}} + p^{-\frac{1+z}{2(1-z)}}}{p^{\frac{1+z}{2(1-z)}} - p^{-\frac{1+z}{2(1-z)}}}}\\ 
\qquad= 1 - \displaystyle{\frac{z}{2\ln p} \frac{d}{dz}} \ln\left(p^{\frac{1+z}{2(1-z)}} - p^{-\frac{1+z}{2(1-z)}}\right), 
\end{array}
& \text{for } \left|z\right|<1\\
 {} & {}\\
 \begin{array}{l}
\,- \displaystyle{\frac{z}{2(1-z)^{2}}\, \frac{p^{\frac{1+z}{2(1-z)}} + p^{-\frac{1+z}{2(1-z)}}}{p^{\frac{1+z}{2(1-z)}} - p^{-\frac{1+z}{2(1-z)}}} }\\
\qquad=  - \displaystyle{\frac{z}{2\ln p} \frac{d}{dz}} \ln\left(p^{\frac{1+z}{2(1-z)}} - p^{-\frac{1+z}{2(1-z)}}\right), 
\end{array}
& \text{for } \left|z\right|>1 
\end{cases}
\label{pZetaRZed}
\end{equation}
Let us reiterate that we do not `derive' the above, but arrive at the expression by demanding that the resolvent satisfies the all 
necessary properties. The coefficients 
\begin{equation}
\beta_n^{(p)} = \frac{1}{2\pi i} \oint \frac{dz}{z^{n+1}} R_<^{(p)}(z)
\end{equation}
(definining the no-gap phase of the matrix model) may in principle be determined from the above with $R^{(p)}_<(z)$ referring to the 
resolvent in the region $|z| < 1$. We shall return to it in \cref{ssec:SimpLocal}.

\subsection{Eigenvalue density from the resolvent}\label{ssec:EValResol}
The eigenvalue density can be obtained from real part of the resolvent \cite{2015arXiv151004430E}
\begin{equation}
\label{eq:rhodef}
2\pi \rho(\theta)  
= \lim_{\epsilon\to 0} \left( R\left((1-\epsilon) e^{i\theta}\right) - R\left((1+\epsilon) e^{i\theta}\right) \right) 
\end{equation}
Alternatively, $2\pi \rho(\theta) = 2\,\text{Re} \left(R(e^{i\theta})\right) - 1$, which relates the density to the real part of the discontinuity 
of the resolvent function across the unit circle. The two definitions are equivalent since $R(z)$  is not really defined on the unit circle 
on the $z$-plane. The imaginary part of $R(z)$, on the other hand, gives the derivative of the potential  \cite{2015arXiv151004430E}  
\begin{equation} \label{eq:V'def2}
V'(\theta) = 2\,\text{Im} \left(R(e^{i\theta})\right)
\end{equation}
that appears on the RHS of the saddle point equation \cref{eq:eveqnplaq}.

Thus the eigenvalue density function and the potential for the matrix model of the local zeta function are determined
from the resolvent
\begin{align}
\begin{split}
2\pi\rho^{(p)}(\theta) 
&=\, 1 + \frac{\pi}{2\ln p\, \sin^2 \frac{\theta}{2}}\, \sum_{n\in\mathbb{Z}} \left[\delta\!\left(\cot \frac{\theta}{2} - 
\frac{2\pi  n}{\ln p}\right) -\, \delta\!\left(\cot\frac{\theta}{2}\right)\right]\\
\frac{dV^{(p)}(\theta)}{d\theta}  
&=\, -\, \frac{1}{2\ln p\, \sin^2 \frac{\theta}{2}}\, \sum_{n\in\mathbb{Z}} \displaystyle\frac{1}{ \cot\frac{\theta}{2} - \frac{2\pi n}{\ln p}}
\end{split}  \label{pRhoV}  
\end{align}	
on the unit circle. If we combine the terms corresponding to positive and negative integers in the second sum above, we can write
\begin{equation*}
\frac{dV^{(p)}(\theta)}{d\theta}  = \frac{1}{4\sin^2\frac{\theta}{2}} \cot\left( \frac{1}{2}\ln p\, \cot\frac{\theta}{2} \right) 
\end{equation*}
Indeed this is the imaginary part of the resolvent calculated from \cref{eq:V'def2}. These functions extremise the effective action
\begin{align}
\begin{split}
S_{\mathrm{eff}}^{(p)} &= - \frac{1}{2}\int d\theta\, \Xint- d\theta'\, \rho^{(p)}(\theta)\, \rho^{(p)}(\theta')\,
\ln\left|4\sin^{2}\left( \frac{\theta-\theta'}{2} \right)\right| \\
&\qquad\qquad - \sum_{n=1}^{\infty}\frac{2}{n}\int d\theta \int d\theta' 
V^{(p)\prime}(\theta)\,\rho^{(p)}(\theta')\, \sin n\theta\, \cos n\theta'
\end{split}
\label{pSeff}
\end{align}
We shall work with this definition of the effective action of the matrix model for the local zeta function at the prime $p$. 

It would appropriate to comment on the eigenvalue distribution \cref{pRhoV} obtained from our proposed resolvent. At the point 
$\theta=0$, the  function $\rho(\theta)$ becomes negative, contrary to what is expected of a density, although it is positive at all 
other points on the unit circle. It is an artefact of the conformal mapping from the $s$- to the $z$-plane. The number of poles of 
the local zeta functions (similarly, the number of zeroes of the adelic or symmetric zeta function) increases as $\mathrm{Im}(s)
\rightarrow\pm\infty$ with $\mathrm{Re}(s)$ fixed to 0 (or ${1}/{2}$ respectively). Consequently, the number of discretely located 
peaks in the eigenvalue density increases as we approach $\theta=0$ on the unit circle in $z$-plane. It is possible to remove the 
point $\theta=0$ from the domain of definition of $\rho(\theta)$, but this will be at the cost of the normalisibility condition on 
$\rho(\theta)$, which in turn is related to $R(0)=1$. This is a technical issue which is unavoidable when accumulation points 
$\mathrm{Re}(s)\pm i\infty$ are brought to a point on the unit circle by a conformal map. In \cref{app:NormResolvent} we show 
that the fluctuations in the density correctly add to zero. A better understanding may be possible if one can develop a phase 
space picture for Hermitian matrix models. However, if we accept it as such, the matrix model technique for unitary ensembles 
will allow us a phase space description. The resolvent is a valid solution to the Hilbert transform arising from extremising the 
action. Therefore, it can be used to calculate the moments of the specific model.    

\section{Matrix model in the phase space}\label{sec:PhaseSpace}
The partition function (and the effective action) of the unitary matrix model can be expressed as an integral over a phase 
space, in which the angular coordinates $\theta$ are augmented by their conjuguate momenta $h$. The objective is to 
write the partition function (schematically) as $Z \sim \int \mathcal{D}\theta\, \mathcal{D}h\, e^{- H(\theta, h)}$, where $H$ 
is a Hamiltonian. We shall review this phase space description, developed in Refs.\cite{Dutta:2007ws,Dutta2016byx} with 
the motivation that phase space picture may suggest a natural Hamiltonian. 

The starting point is the partition function that can alternatively be expressed as a sum over representations of $U(N)$, 
which are characterized by Young diagrams. In the large $N$ limit, the contributions from a class of diagrams 
dominate\cite{Douglas:1993iia}. It is possible to express the partition function in terms of the phase space of a system of 
free fermions. In this picture the coordinates and momenta are the eigenvalues $\theta_i$ and the number of hooks $h_i$
of the right-most boxes in $i$-th row of a Young diagram. 

At this point let us pause for a moment to summarise (purging some technical details for simplicity) the approach we adopt 
towards constructing a Hamiltonian related to the non-trivial zeroes of the Riemann zeta function. We will use the recently
developed phase space formalism of unitary matrix model.
\begin{enumerate}
\item 
First, we construct unitary matrix models corresponding to each prime number starting with the resolvent \cref{pZetaRZed} by 
mapping the imaginary axis (containing the simple poles of the local zeta function) to the unit circle. We will refer to these as 
the \emph{local models}.
\item 
Secondly, we rewrite the partition function of these matrix models in the schematic form 
\begin{equation*}
Z_{(p)} \sim \int \mathcal{D}\theta_{(p)}\, \mathcal{D}h_{(p)}\, e^{- H_{(p)} (\theta_{(p)},h_{(p)})}
\end{equation*}
We then realise the partition function $Z_{(p)}$ as the trace of an operator to extract the operator form of the Hamiltonian. It should 
be mentioned that in the classical (large $N$) limit, the operator formalism is in the Koopman-von Neumann formalism of classical 
mechanics. Also that results from $p$-adic analysis help us in realising this step.
\item 
Next we combine (the partition functions of ) all the \emph{local models} to form a matrix model for the Riemann zeta function. The
parameters of this matrix model is divergent, therefore, we discuss ways of regularisation to make it well defined.   
\item
Finally, we discuss the construction of Wigner functions in the phase space of the matrix models. This will lead to a connection to
the prime counting function. 
\end{enumerate}	
In the rest of the paper, after recalling some relevant aspects of the phase space of the UMM, we will provide details of these steps 
and present the results that follow.
\subsection{Sum over representations}\label{ssec:SumRep}
The exponential in the partition function \cref{PFPlaq} of a single plaquette model may be expanded as
\begin{align}
\begin{split}
Z &= \int \mathcal{D}U\, \left(\sum_{\mathbf{k}} \frac{\varepsilon(\boldsymbol{\beta},\mathbf{k})}
{\Delta({\mathbf{k}})} \Upsilon_{\mathbf{k}}(U)\right)\, \left(\sum_{\mathbf{l}} \frac{\varepsilon(
\boldsymbol{\beta},\mathbf{l})} {\Delta({\mathbf{l}})} \Upsilon_{\mathbf{l}}(U^{\dagger})\right)\\
\text{where,}\;
\varepsilon(\boldsymbol{\beta}, \mathbf{k}) &= \prod_{n=1}^{\infty}N^{k_n}\beta_{n}^{k_n}, \quad
\Delta({\mathbf{k}}) = \prod_{n=1}^{\infty} k_{n}! \, n^{k_n}\: \text{ and } \: 
\Upsilon_{\mathbf{k}}(U)=\prod_n (\mathrm{Tr}\, U^n)^{k_n}
\end{split}
\end{align}
Now $\Upsilon_{\mathbf{k}}(U)$ can be expanded as
\begin{equation}
\Upsilon_{\mathbf{k}}(U) = \sum_{R\in\mathrm{irreps}} \chi_{R}(C(\mathbf{k})) \mathrm{Tr}_{R}[U] 
\end{equation}
where $\chi_{R}(C(\mathbf{k}))$ is the character of the conjugacy class $C(\mathbf{k})$ of the permutation group 
$S_{K}$ ($K=\sum_n nk_n$) and $R$ denotes an irreducible representation of $U(N)$. Finally, using the orthogonality 
of the characters
\begin{equation}
\int {\mathcal{D}}U\, \mathrm{Tr}_{R}[U]\, \mathrm{Tr}_{R'}[U^{\dagger}] = \delta_{RR'}
\end{equation}
one obtains
\begin{equation}
Z = \sum_{R\in\mathrm{irreps}} \sum_{\mathbf{k}} \frac{\varepsilon(\boldsymbol{\beta}, \mathbf{k})}{\Delta(\mathbf{k})} 
\sum_{\mathbf{l}} \frac{\varepsilon(\boldsymbol{\beta},\mathbf{l})}{\Delta({\mathbf{l}})} \chi_{R}(C(\mathbf{k}))
\chi_{R}(C(\mathbf{l}))
\end{equation}
as the partition function.

The sum over representations of U($N$) can be traded for a sum over the Young diagrams, with a maximum of $N$ rows 
and an arbitrary numbers of boxes in each row, as long as the number of boxes in a particular row does not exceed that in 
a row preceding it. Let $\lambda_j$ be the number of boxes in $j$-th row, and $\sum_j \lambda_j = K$. 
\begin{equation}\label{eq:pffinal}
Z = \sum_{\boldsymbol{\lambda}} \sum_{\mathbf{k}, \mathbf{l}} 
\frac{\varepsilon(\boldsymbol{\beta}, \mathbf{k})\varepsilon(\boldsymbol{\beta}, 
\mathbf{l})}{\Delta({\mathbf{k}}) \Delta({\mathbf{l}})} \chi_{\mathbf{\lambda}}\left(C
(\mathbf{k})\right) \chi_{\mathbf{\lambda}}\left(C(\mathbf{l})\right)
\delta\left(\sum_n n k_n - \sum_j \lambda_j\right) \delta\left( \sum_n n l_n - 
\sum_j \lambda_j\right)
\end{equation}
Note that the total number of boxes is the same as the order of the permutation group $S_K$.  The characters of the 
conjugacy classes are determined recursively by the Frobenius formula (see, e.g. \cite{fulton-harris,hamermesh,lassalle}). 

In the large limit $N\to\infty$, we introduce the continuous variables
\begin{equation}
x = \frac{i}{N}\in [0,1] \qquad \text{and}\qquad h_i \rightarrow N h(x)
\end{equation}
(where the decreasing sequence of numbers $h_i = \lambda_i + N - i$ denote the hook length) in terms of which the 
partition function, in the large $N$ limit, is
\begin{equation}\label{eq:pfyt1}
Z = \int \mathcal{D} h(x)\, \prod_n \int d\mathbf{k}' d\mathbf{l}' \exp\left( -N^2 
\widetilde{S}_{\mathrm{eff}} \left[ h(x),\mathbf{k}', \mathbf{l}' \right]\right)
\end{equation}
where $\widetilde{S}_{\mathrm{eff}}$ is an effective action defined in terms of $h(x)$ and the parameters 
$\mathbf{k} = N^2\mathbf{k}'$. The dominant contribution at large $N$ comes from those representations which extremize 
$\widetilde{S}_{\mathrm{eff}}$. These representations are characterised by the so called \emph{Young diagram density} 
\begin{equation}
u(h) = - \frac{\partial x}{\partial h}
\end{equation}
The fact that $h(x)$ is monotonically decreasing, ensures that $0<u(h)\le 1$ for all $x\in[0,1]$, as required of a density.

In the generic unitary one plaquette model, it is difficult to find $u(h)$ by extremizing the effective action $\widetilde{S}$. 
This is due to the fact that explicit expressions for the characters of a permutation group in the presence of non-trivial 
cycles are not available.

The character is determined from the Frobenius formula as
\begin{equation}
\chi_{\mathbf{h}}\left(C(\mathbf{k})\right) = \left\{ \left(\sum_{i=1}^{N} x_{i}\right)^{k_{1}}
\left(\sum_{i=1}^{N} x_{i}^{2}\right)^{k_{2}} \cdots\, \prod_{i<j}(x_{i}-x_{j})\right\}_{\mathbf{h}=(h_{1},..h_{N})}
\end{equation}
where $\left\{g(\mathbf{x})\right\}_{\mathbf{h}=(h_{1},..h_{N})}$ denotes the coefficient of the term $x_1^{h_1}\cdots 
x_N^{h_N}$ in $g(\mathbf{x})$. In order to select these coefficients, let us introduce $N$ complex variables $(z_1, z_2, 
\cdots z_N)$ to write the character as an integral
\begin{equation}\label{eq:character2}
\chi_{\mathbf{h}}\left(C_{\mathbf{k}}\right) = \prod_{i}^{N} \oint\frac{dz_i}{2\pi i}
\frac{1}{z_{i}^{h_{i}+1}} \left[ \left(\sum_{i=1}^{N}z_{i}\right)^{k_{1}}
\left(\sum_{i=1}^{N} z_{i}^{2}\right)^{k_{2}} \cdots\, \prod_{i<j}(z_{i}-z_{j})\right]
\end{equation}
where the contours are taken around the origin. We should mention that in the above, $z_i$ are auxiliary variables, 
introduced as variables of integration. The notation, however, is chosen in anticipation of the fact that they will soon be 
identified with the eigenvalues of the U($N$) matrices. 

\subsection{Large $N$ saddle}\label{ssec:NSaddle}
We now exponentiate the terms $\left(\sum z_i\right)^{k_i}$, take the large $N$ limit where we use the continuous functions 
\begin{equation}
\frac{z_i}{N} \rightarrow z(x),\qquad \frac{h_i}{N} \rightarrow h(x), \quad \text{and}\quad 
k_{n} \rightarrow N^{2}k_n'
\end{equation}
of $x={i}/{N}$, and replace the sum over $i$ by an integral over $x$, to rewrite the characters as
\begin{align}\label{eq:character1}
\begin{split}
\chi\left(h(x), C(\mathbf{k}')\right) &= \frac{1}{(2\pi i)^N} \oint \frac{\mathcal{D}z(x)}{z(x)}
\exp\left( - N^2 S_{\chi}\left[h(x),\mathbf{k}'\right]\right)\\
S_{\chi}\left[h(x),\mathbf{k}'\right] &= - \sum_n k_n' \left[ (n+1) \ln N + \ln \int_0^1 dx\, 
z^n(x) \right] + \int_0^1 dx\, h(x) \ln z(x)\\
&\qquad\qquad - \frac{1}{2} \int_0^1 dx\, \Xint -_0^1 dy\, \ln \left| z(x)-z(y) \right|  + 
\frac{1}{N^2} K \ln N
\end{split}
\end{align}
In spite of the notation, $S_\chi \left[h(x),\mathbf{k}'\right]$ is not really an action. 

In the large $N$ limit, the dominant contribution to the characters arise from a saddle point that correspond to a particular 
distribution of $z(x)$. The conditions to determine the extremum of $S_{\chi}$ are
\begin{align}
\begin{split}
\sum_n n k_n' \frac{z^n(x)}{ \int_0^1 dx' z^n(x') }  + \Xint-_0^1 dy \frac{z(x)}{z(x)-z(y)} - h(x) &= 0\\
\sum_n n k_n' \frac{z^n}{\int dz' z^{'n}\rho(z')} + \Xint- dz' \frac{z \rho(z')}{z - z'} - h(z) &= 0
\end{split}
\end{align}
where we have introduced $\rho(z) = {\partial x}/{\partial z}$ in the second line. Since the contours in \cref{eq:character2})
go around the origin, we can take $z=e^{i\theta}$ and write the above as
\begin{equation}\label{eq:chaeqn}
\sum_nn k_n' \frac{e^{i n \theta}}{\int d\theta' \rho(\theta') e^{in\theta'}} + \Xint-_{-\pi}^{\pi} 
d\theta' \rho(\theta') \frac{e^{i\theta}}{e^{i\theta}-e^{i\theta'} } - h(\theta) = 0
\end{equation}
where now, with an abuse of notation, $\rho(\theta) = \frac{\partial x}{\partial\theta} = i e^{i\theta} \rho(z)$ 
naturally. The imaginary part of the above
\begin{equation}
\sum_n n k_n' \frac{\sin n\theta}{\int d\theta' \rho(\theta') \cos{n\theta'}} - \frac{1}{2} \Xint- d\theta' 
\rho(\theta') \cot{\frac{\theta - \theta'}{2}} = 0
\end{equation}
is exactly the same as the eigenvalue equation \cref{eq:eveqnplaq} with $n k'_n = \beta_n\,\left(\int d\theta' \rho(\theta') 
e^{in\theta'}\right)$, which is the equation of motion for ${\mathbf{k}}'$. This identification is actually consistent and the 
auxiliary variables which appear in the Frobenius formula can indeed be thought of as the eigenvalues of the unitary 
matrices under consideration. Not only do they satisfy the same equation as the eigenvalues, the partition function written 
in terms of these variables exactly matches with the partition function written in terms of the eigenvalues. Thus, in the large 
$N$ limit, the numbers of cycles $k_n'$ are fixed in terms of the parameters $\beta_n$. The density $\rho(\theta)$ above is 
the same as the density of eigenvalues. For more details, we refer to \cite{Dutta2016byx}.

The real part of \cref{eq:chaeqn}
\begin{equation}
h(\theta) = \frac{1}{2} + \sum_n \beta_n \cos{n\theta}
\end{equation}
is an algebraic equation that relates the number of boxes in a Young diagram to the eigenvalues of 
unitary matrices. 

In terms of $\theta$ and $\rho(\theta)$, $S_\chi$ can be expressed as 
\begin{align}\label{eq:Schi2}
\begin{split}
- S_{\chi} \left[h(x),\mathbf{k}'\right] &= \sum_n k_n' \left( (n+1)\ln N + \ln \left( \int d\theta \rho(\theta) 
\cos n\theta\right)\right)  - K' \ln N\\
&\qquad+ \frac{1}{2} \int d\theta \rho(\theta)\Xint - d\theta' \rho(\theta') 
\ln\left| e^{i\theta} - e^{i\theta'} \right| - i \int d\theta \rho(\theta)h(\theta) \theta
\end{split}
\end{align}
Since $\rho(\theta)$ is an even function, the last terms implies a redundancy in $h(\theta)$. Different Young distributions 
which are related by $S_\chi\left[ h(\theta),\mathbf{k}\right] = S_\chi\left[ h(\theta) + f(\theta),\mathbf{k}\right]$, for any 
even function $f(\theta)$, yield the same eigenvalue distribution $\rho(\theta)$. Hence, the most general relation between 
$h(\theta)$ and $\theta$ is
\begin{equation}\label{eq:generichtheta}
h(\theta) = \frac{1}{2} + \sum_n \beta_n \cos{n\theta} + f_{\mathrm{even}}(\theta)
\end{equation}
where $f_{\mathrm{even}}(\theta)$ is an even function of $\theta$. This redundancy plays an important role in determining 
different phases of a unitary matrix model. In Ref.\cite{Dutta2016byx} the equation above was called the \emph{boundary 
equation} as it defines the boundary of the density distribution in phase space. However, we shall postpone an analysis of 
this issue in the context of the matrix model of the local zeta function to the future.

\subsection{Many-body wave function}\label{ssec:WaveFn}
The results in \cref{ssec:SumRep,ssec:NSaddle}, in particular \cref{eq:pffinal,eq:pfyt1,eq:character1}, allows one 
to write the partition function of a unitary matrix model as
\begin{align}
\begin{split}
Z &= \oint \frac{\mathcal{D}z(x)}{z(x)} \oint \frac{\mathcal{D}z'(x)}{z'(x)} \int \mathcal{D}h(x)\, 
\exp\left(-N^2 S_{\chi}\left[z,h\right]\right) \, \exp\left(-N^2 S_{\chi}\left[z',h\right]\right)\\
S_{\chi} \left[z,h\right] &= \int dx\, \left(- \sum_{n=1}^{\infty} \frac{1}{n}\beta_n z^n(x) + h(x)\ln z(x) -
\frac{1}{2}\Xint-dy \ln \left| z(x) - z(y)\right| \right)  
\end{split}
\end{align}
The above is related to the wavefunction of $N$ fermions 
\begin{align}
\begin{split}
e^{- N^2 S_{\chi}}(z_{i},h_{i}) &= \left(\prod_{i<j} (z_{i}-z_{j})\right) \, \exp\left(\sum_{i=1}^N \left[ 
N \sum_{n=1}^{\infty} \frac{\beta_n}{n} z^n_i - h_i \ln z_i\right]\right)\\
&= \det \left[ z_i^{j - 1 - h_i} \exp\left( N \sum_{n=1}^\infty \frac{\beta_n}{n} z^n_i\right) \right]
\end{split}
\end{align}
The many body wavefunction $\boldsymbol{\Psi}(\mathbf{z},\mathbf{h}) = e^{-N^2S_\chi}$ leads to the 
`fermion density' function $\boldsymbol{\Omega}(\mathbf{z},\mathbf{h}) = \boldsymbol{\Psi}^\dagger 
\boldsymbol{\Psi}$.

\subsection{Simplifying the local model}\label{ssec:SimpLocal}
Let us try to simplify some of the expressions. The potential term in the matrix model involves an infinite sum 
$\sum_n\beta_n \mathrm{Tr}(U^n)$ (and its conjugate). Using the fact that the coefficients $\beta_n$ were 
determined from the resolvent, and the expression \cref{pZetaRZed} for the latter, we can perform the sum  
\begin{align}\label{eq:actionp}
\begin{split}
\sum_{n=1}^\infty\frac{\beta^{(p)}_n z_i^n}{n}  &= -\, \frac{1}{4\pi i \ln p} \sum_{n=1}^\infty \frac{z_i^n}{n} 
\oint\frac{dz}{z^{n}}\, \frac{d}{dz}\ln\left(p^{\frac{1+z}{2(1-z)}} - p^{-\frac{1+z}{2(1-z)}}\right)\\
&= \frac{1}{4\pi i \ln p} \oint dz\, \ln\left(1 - \frac{z_i}{z}\right)\, \frac{d}{dz}\ln\left(p^{\frac{1+z}{2(1-z)}} - 
p^{-\frac{1+z}{2(1-z)}}\right)\\
&= -\, \frac{1}{2\ln p} \ln\left(\frac{p^{\frac{1+z_i}{2(1-z_i)}} - p^{-\frac{1+z_i}{2(1-z_i)}}}{p^{1/2}-p^{-1/2}}\right)\\ 
&= -\, \frac{1}{2\ln p} \ln\left(\frac{p^{\frac{i}{2}\cot\frac{\theta_i}{2}} - p^{-\frac{i}{2}\cot\frac{\theta_i}{2}}}{p^{1/2}-
p^{-1/2}}\right)  
\end{split}
\end{align}
where we have substituted $z_i=e^{i\theta_i}$ in the last line. With this we can write 
\begin{align}
\prod_{i=1}^N \int \frac{dz_i}{z_i} e^{-N^2 S^{(p)}_\chi} = & \prod_{i=1}^{N}\int \frac{2\,d\!\left(\cot
\frac{\theta_{i}}{2}\right)\, e^{-ih_{i}\theta_{i}}}{(1+i\cot\frac{\theta_{i}}{2})^{N}(1-i\cot\frac{\theta_{i}}{2})}\,
\prod_{i<j}2\left(\cot\frac{\theta_{i}}{2}-\cot\frac{\theta_{j}}{2}\right)\nonumber\\
&\qquad\times\,
\prod_{i=1}^{N} \left(\frac{(1-p^{i\cot\frac{\theta_{i}}{2}})(1-p^{-i\cot\frac{\theta_{i}}{2}})}{(p^{1/2}-
p^{-1/2})^2}\right)^{-\frac{N}{2\ln p}}\label{eq:partitionp}\\
Z_{(p)} = \prod_{i=1}^{N} \int \frac{2\,d\!\left(\cot\frac{\theta_{i}}{2}\right)}{(1+\cot^{2}\frac{\theta_{i}}{2})^{N}}\,
&\prod_{i<j}4\left(\cot\frac{\theta_{i}}{2}-\cot\frac{\theta_{j}}{2}\right)^{2}\,
\prod_{i=1}^{N}\left[\frac{(p^{1/2}-p^{-1/2})^2}{2 - p^{i\cot\frac{\theta_{i}}{2}} - p^{-i\cot\frac{\theta_{i}}{2}}}
\right]^{\frac{N}{2\ln p}}\nonumber
\end{align}
where, to get to the partition function in the last line, we have performed the integrals over the momenta $h_{i}$, which
yielded $\delta(\theta_{i}+\theta'_{i})$, using which we eliminated one set of integrals over the coordinates $\theta$.
Schematically, the partition function of the matrix model is of the form of the local zeta-function.

\subsection{Density function in the phase space}\label{ssec:PhSpDensity}
Both the eigenvalue distribution and the distribution of the Young diagram (of at least a particular class of unitary matrix 
models) can be obtained from a single distribution function \cite{Dutta:2007ws,Dutta2016byx}. This function, which is one 
in some region of the $(\theta,h)$ space, and zero outside it, may be identified with the phase space distribution of $N$ 
free fermions in one dimension. 

Let us define a complex phase space distribution function $\Omega(\theta,h)$ in the $(\theta,h)$ space such that
\begin{align}
\begin{split}\label{eq:gen-phasespace}
\int_0^\infty \Omega(\theta,h) dh &= 2R(e^{i\theta}) - 1 = \varrho(\theta) \\
\int_{-\pi}^{\pi} \Omega(\theta,h) d\theta &= u(h)
\end{split}
\end{align}
When integrated over $h$, the real part of $\Omega$ gives the eigenvalue density $\rho(\theta)$, while the imaginary part gives the 
derivative of the potential $V'(\theta)$. Since these are even and odd functions of $\theta$, respectively, the real and imaginary part 
of $\Omega(\theta,h)$ are also, respectively, even and odd. Hence, when we integrate $\Omega(\theta,h)$ over $\theta$ only the real 
part contributes to give the Young diagram distribution function $u(h)$ above. Although  the distribution function is simple, the shape 
(or topology) of the imaginary part of $\Omega(\theta,h)$, which contains information about the derivative of the potential, could be 
complicated, but is fixed for a given matrix model. Whereas the topology of its real part will change as we go from one phase to 
another.
 
\subsection{Phase space for the local model}\label{ssec:PhSpLocZ}
We have seen that in the no-gapped phase, $2\pi\rho^{(p)}(\theta) = 1 + \sum_n\beta^{(p)}_n\cos n\theta$ (where the coefficients 
$\beta_p^{(p)}$ are to be determined from the resolvent function \cref{pZetaRZed}) gives the density of the eigenvalues on the unit 
circle of the matrix model for the local zeta function. Together with $V^{(p)\prime}(\theta) = \sum_n\beta^{(p)}_n\sin n\theta$  one can 
define a complexified eigenvalue density $\varrho^{(p)}(\theta) = 2 R^{(p)}(\theta) -1 = 2\pi\rho^{(p)}(\theta) + iV^{(p)\prime}(\theta)$. 
This function carries the information of the poles of \cref{pZeta}, the local  zeta function. 

Recall the resolvent for the local zeta function proposed in \cref{pZetaRZed}. This gives 
\begin{equation}
\varrho^{(p)}(\theta) = 1 + \frac{1}{4\sin^2\frac{\theta}{2}} \,\frac{1 + p^{-i\cot\frac{\theta}{2}}}{1 - p^{-i\cot\frac{\theta}{2}}} 
= 1+ \frac{1}{4\sin^2\frac{\theta}{2}}\left(1+ 2\sum_{n=1}^\infty p^{-i n \cot\frac{\theta}{2}}\right) 
\end{equation}
where the ability to express it as a sum suggests that we may be able to make sense of it as the trace of some operator. We shall now 
argue that this is indeed possible, albeit not on a Hilbert space that is very familiar to physicists. Rather the operator is on a 
\emph{subspace} of $L^2\left(\Qp\right)$, the space of square integrable complex valued functions on the $p$-adic line. In fact a 
comparison of the above with \cref{VDonKozy} makes it immediately obvious that the operator is related to the (generalized) Vladimirov 
derivative $D^{-i\cot\frac{\theta}{2}}$ defined in \cref{VladD}. We write
\begin{align}
\begin{split}
\varrho^{(p)}(\theta) &= 1 + \frac{1}{4\sin^2\frac{\theta}{2}} + \frac{1}{2\sin^2\frac{\theta}{2}} \,
\mathrm{Tr}_{\mathcal{H}_-^{(p)}} \left(D_{(p)}^{-i\cot\frac{\theta}{2}}\right)\\
& =  1 + \frac{1}{4\sin^2\frac{\theta}{2}} + \frac{1}{2\sin^2\frac{\theta}{2}} \,\sum_{n=1}^\infty \left\langle 
\psi^{(p)}_{-n+1,0,1} \left| D_{(p)}^{-i\cot\frac{\theta}{2}} \right| \psi^{(p)}_{-n+1,0,1} \right\rangle   
\end{split}\label{RhopAsTrace}
\end{align}
where the trace is expressed in terms of the (normalised) Kozyrev wavelets  $\left| \psi^{(p)}_{-n+1,0,1}\right\rangle$ defined in 
\cref{pWavelet} for $n\in\mathbb{N}$. With this choice, the translation parameter $m$ gets restricted to a finite set, as discussed at 
the end of \cref{sec:Zeta}. However, we can consistently restrict to the subspace in which $m=0$. The parameter $j$ may also be 
restricted to unity without any loss of generality. The set of Kozyrev wavelet functions, which are eigenfunctions of the Vladimirov 
derivative form a basis for complex valued, mean-zero, square integrable functions on the $p$-adic space $\Qp$ \cite{Kozyrev:2001}. 
Our restricted set of functions span a subspace $\mathcal{H}_-^{(p)}\subset L^2\left(\Zp\right)\subset L^2\left(\Qp\right)$ of complex 
valued, mean-zero, square integrable functions supported on the compact set $\Zp \subset \Qp$. 

It should be emphasised that notwithstanding the notation, we are dealing with a classical system, therefore, the states are classical 
states. This is in the sense of Koopman-von Neuman classical mechanics which uses the Hilbert space formalism\cite{wilczek,kvnthesis}. 
As in quantum mechanics, classical states are vectors in a Hilbert space and physical observables are Hermitian operators, however, 
unlike in quantum theory, the operators commute. The dynamical equation is the Liouville equation satisfied by the `wavefunction'. For 
further details, we refer to the articles cited above.

We would like to arrive at the density $\varrho^{(p)}(\theta)$ from a density in phase space. This is possible thanks to the integral 
\cref{eq:pInt}
\begin{align}
\begin{split}
\varrho^{(p)}(\theta) &= 1+ \frac{1}{4\sin^{2}\frac{\theta}{2}} + \frac{1}{2\sin^2\frac{\theta}{2}}\, \frac{p}{(p-1)}\,
\int_{p\Zp} |\mathfrak{h}|_p^{i\cot\frac{\theta}{2} - 1} d\mathfrak{h} \\
& = p\int_{p\Zp} d\mathfrak{h} \left(1 + \frac{\csc^{2}\frac{\theta}{2}}{4} + \frac{\csc^{2}\frac{\theta}{2}}{2(p-1)}\, 
|\mathfrak{h}|_p^{i\cot\frac{\theta}{2}-1}\right)
\end{split}
\end{align}
from which we can read the phase space density $\Omega^{(p)}(\theta,h)$
\begin{equation}\label{locPhSpRho}
\Omega^{(p)}(\theta,\mathfrak{h}) = p \left(1 + \frac{\csc^{2}\frac{\theta}{2}}{4} + \frac{\csc^{2}\frac{\theta}{2}}{2(p-1)}\, 
|\mathfrak{h}|_p^{i\cot\frac{\theta}{2}-1}\right) 
\end{equation}
where $\theta\in\left[0,2\pi\right]\subset\mathbb{R}$ and $\mathfrak{h}$ takes values in $\mathbb{Z} \subset \Zp\subset\Qp$. The 
momentum density function $u^{(p)}(\mathfrak{h})$ \cref{eq:gen-phasespace} is obtained by integrating $\Omega^{(p)}(\theta,
\mathfrak{h})$ over the eigenvalues\footnote{In evaluating the integral the contour is chosen so that only the pole at $z=0$ contributes. 
The singularities on the unit circle $|z|=1$ are avoided, including the essential singularity at $z=1$. The latter is related to the problem of 
the positivity of the density function while mapping the imaginary line to the unit circle. See \cref{app:NormResolvent}.} $z=e^{i\theta}$
\begin{equation}
u^{(p)}(\mathfrak{h}) = -ip \oint \frac{dz}{z}\left(1- \frac{z}{(1-z)^{2}} - \frac{2z}{(p-1)(1-z)^{2}} 
|\mathfrak{h}|_p^{\frac{1+z}{1-z}-1}\right) \label{pmomspacedensity}
\end{equation}
Evaluating the integral we find that the momentum distribution function is a constant $2\pi p$, which means that the momenta are 
equally spaced. This is consistent with the spectrum of the operator $D^\alpha$ (with $\alpha = - i\cot\frac{\theta}{2}$) in 
\cref{VDonKozy}.

\subsection{Hamiltonian for the local model}\label{ssec:LocHamiltonian}
The fluctuating part of the phase space density \cref{locPhSpRho}, without the measure factor, is
\begin{equation}
\delta\Omega^{(p)} = D_{(p)}^{-i\cot\frac{\theta}{2}} \sim p^{-in\cot\frac{\theta}{2}} = 
\exp\left({-in\ln p\, \cot\frac{\theta}{2}}\right)
\end{equation}
This satisfies the Liouville equation
\begin{equation}\label{eq:Liouville}
\left(\frac{\partial H^{(p)}(X_{(p)},P_{(p)})}{\partial X_{(p)}}\frac{\partial}{\partial P_{(p)}} - 
\frac{\partial H^{(p)}(X_{(p)},P_{(p)})}{\partial P_{(p)}}\frac{\partial}{\partial X_{(p)}}\right) \,
\delta{\Omega}^{(p)} (X_{(p)},P_{(p)}) = 0
\end{equation}
for the Hamiltonian $H^{(p)}=n\ln p\cot\frac{\theta}{2}$, which is a function of generalized coordinate $X_{(p)}=
\cot\frac{\theta}{2}$ and momentum $P_{(p)}\sim\ln D_{(p)} \sim n\ln p$. We see that $\delta\Omega^{(p)} =
e^{-iH^{(p)}}$, although in reality any function of the Hamiltonian will satisfy the Liouville equation for the density.  

This Hamiltonian is of the form $H=xp$ as prescribed in Berry and Keating \cite{BK,BerryKeating2}, but expressed in 
the Koopman-von Neumann operator formalism for a classical dynamical system. Moreover, the metric on the integer 
valued momenta is the non-archimedean $p$-adic metric. The Berry-Keating Hamiltonian has a continuous spectrum 
and is integrable, while any Hamiltonian related to the Riemann zeta function is expected to be chaotic. Furthermore, 
the spectrum of the corresponding quantum system is not discrete. In order to address these problems, the authors of 
Ref.\cite{BerryKeating2} suggested imposing restrictions on the accessible region of the phase space. It is interesting 
to note that for the trace in \cref{RhopAsTrace} to be meaningful, that is, for the operator to be trace class, we have a 
natural truncation to the subspace $\mathcal{H}_-^{(p)}$ of $L^2(\Qp)$. The action of the Vladimirov derivative 
$D_{(p)}$ naturally restricted to this subspace. In fact, the subspace is defined by the span of a subset of orthonormal 
eigenvectors of this operator.
 
\section{Matrix model in the large phase space}\label{sec:Allp}
We shall now attempt to construct a matrix model for the Riemann zeta function, by combining the data from the local 
zeta functions for all primes. Naturally this involves taking a product (of the infinite number) of partition functions we 
obtained for the primes individually. Not unexpectedly, some of the results are divergent. We shall adopt the usual 
approach employed by physicists in treating divergences and try to extract a finite answer through a process of 
renormalisation.

\subsection{Combining the local information}\label{ssec:synth}
In order to define the model we need to specify the potential, which in turn is determined by the coefficients 
$\{\beta_m\}$. Naturally $Z=\prod_p Z_p$ and $S_{\mathrm{eff}} = \sum_p S^{(p)}_{\mathrm{eff}}$. Let us, therefore, 
consider the sum these coefficients for the local models. However, to conform to the standard convention, we include 
an additional factor of $\ln p$ in the sum. 
\begin{align}
\begin{split}
\beta_m &= \sum_{p\in\mathrm{primes}} \ln p \,\beta_m^{(p)}
= \sum_{p} \ln p\, \frac{1}{2\pi i} \oint  \frac{dz}{z^{m+1}}\, R^{(p)}_<(z)\\
&= \frac{1}{4\pi} \int^{\infty}_{-\infty} d\!\left(\!\cot\frac{\theta}{2}\right) e^{-im\theta} \sum_{p} \ln p\, \left(\sum_{n=1}^\infty 
\left\langle \psi_{-n+1,0,0}^{(p)} \left| D^{-i\cot\frac{\theta}{2}}_{(p)}\right| \psi_{-n+1,0,0}^{(p)} \right\rangle + \frac{1}{2}\right)\\
&= \frac{1}{4\pi} \int^{\infty}_{-\infty} d\!\left(\!\cot\frac{\theta}{2}\right) e^{-im\theta}\, \left(
\mathrm{Tr}_{\mathcal{H}_-}\! \left(\mathbb{D}^{-i\cot\frac{\theta}{2}}\right) + \sum_{p} \ln\sqrt{p} \right) 
\end{split}\label{BigBetaM}
\end{align}
In the following we explain the notation in the first term, as well as the apparently divergent second term, in the last line.

We have defined a \emph{large Hilbert space} 
\begin{equation}
\mathcal{H}_- = \bigotimes_p \mathcal{H}^{(p)}_- = \bigotimes_p \text{Span} \left\{ \sqrt{\ln p}\, 
|\psi^{(p)}_{-n+1,0,1}\rangle\, :\, n\in\mathbb{N}\right\}  \label{lHs}
\end{equation}
in which we have scaled the vectors in the $p$-th Hilbert space by a factor of $\sqrt{\ln p}$. The generalised Vladimirov 
operator $\mathbb{D}^\alpha$ on the large Hilbert space is defined by the equation above. We may take it to be  
\begin{equation}
\mathbb{D}^\alpha = \prod_p D_{(p)}^\alpha 
\label{bigDalpha}
\end{equation}
The eigenvalues of this operator are $p^{n\alpha}$ ($n\in\mathbb{Z}$) and the corresponding eigenvectors 
are $\displaystyle{\Psi_{-n+1,0,1} \equiv \psi^{(2)}_{1,0,1} \otimes \psi^{(3)}_{1,0,1} \otimes \cdots \otimes 
\underbrace{\psi^{(p)}_{-n+1,0,1}}_{p\text{-th place}} \otimes \cdots}$, i.e., it has the wavelet that scales by $p^n$ at the 
$p$-th place, and the basic mother wavelets at all other places. These eigenvectors have an adelic flavour. We need only 
those vectors with $n \in \mathbb{N} \subset \mathbb{Z}$.
  
Now it is not hard to see that 
\begin{equation}
\beta_m = \frac{1}{4\pi}\int_{-\infty}^{\infty} d\!\left(\!\cot\frac{\theta}{2}\right)\, e^{-im\theta}\, \left[   
\int_{0}^{\infty} dx\, x^{-i\cot\frac{\theta}{2}} \,\frac{d\psi(x)}{dx} + \sum_{p} \ln\sqrt{p} \right] \label{betaRev}
\end{equation}
where $\psi(x)$ is the summatory von Mangoldt function defined in \cref{sumVM}. This expression, however, is formal due 
to singularities in the integral over $\theta$. 

Now we turn to the second term, the contribution of which to the partition function is
\begin{equation*}
- \sum_{p}\ln p\, \sum_{n}\frac{1}{n} \oint\frac{dz}{4\pi i} \frac{z_i^n}{z^{n}(1-z)^2} 
= -\frac{z_i}{2 (1-z_i)} \sum_{p}\ln p = \frac{1}{4}\left(1 + i\cot\frac{\theta_i}{2}\right)\sum_p\ln p
\end{equation*}
The partition function also receives a contribution from the hermitian conjugate of this term. When the two are added, the 
divergent $\theta_{i}$ dependent terms cancel pairwise, and we get $\sum_p\ln\sqrt{p}$. This was also at work when we 
went from \cref{eq:actionp} to \cref{eq:partitionp}. Although this factor is infinite, it affects only the overall normalisation of 
the partition function, and not the computation of correlation functions.

\subsection{Renormalization of the parameters}\label{ssec:Renorm} 
Let us try to make sense of the divergent parameters $\beta_m$ by `renormalisation'. To this end, we `renormalise' the 
Kozyrev wavelet states for each prime $p$ by multiplying with a constant. To be precise, let us replace 
$\left| \psi_{-n+1,0,1}^{(p)}\right\rangle \rightarrow c_{n}^{(p)} \left| \psi_{-n+1,0,1}^{(p)}\right\rangle$
so that the \emph{renormalised} parameters $\beta_m^{\mathrm{ren}}$ are
\begin{equation}\label{renBeta}
\beta_m^{\mathrm{ren}} = \frac{1}{4\pi} \int^{\infty}_{-\infty} d\left(\cot\frac{\theta}{2}\right) e^{-im\theta} 
\sum_{p} \ln p\,\sum_{n=1}^\infty \left| c_{n}^{(p)}\right|^2 \left\langle \psi_{-n+1,0,1}^{(p)} \left| 
D^{-i\cot\frac{\theta}{2}}_{(p)}\right| \psi_{-n+1,0,1}^{(p)}\right\rangle
\end{equation}
A convenient choice for the constants is  $c^{(p)}_{n} = p^{-n\mu/2}$ for some real $\mu$. This is equivalent to a shift in 
the exponent of the Valdimirov derivative $D_{(p)}$. Notice that a multiplicative renormalisation of the eigenstates of 
$D_{(p)}$ is equivalent to a redefinition of the operator itself because of the special functional form of its eigenvalues. 
As an aside, we note that this renormalisation is not unique, as one can multiply this choice by an infinite series starting 
with 1. This is equivalent to multiplying the shifted Vladimirov derivative by an infinite series $f\left(D_{(p)}\right) = 1 + a_1 
\left(D_{(p)}\right)^{\mu_1} + a_2 \left(D_{(p)}\right)^{\mu_2} + \cdots$, as long as the additional terms lead to convergent 
integrals. This will of course change the form of the $\beta_m^{\mathrm{ren}}$ from its simplest form above. Therefore, 
we shall use the simplest renormalised form.   

As a result of the renormalisation, the parameters now take the following form 
\begin{align}
\begin{split}\label{renBeta2}
\beta_m^{\mathrm{ren}} &= \frac{1}{4\pi} \int^{\infty}_{-\infty} d\left(\cot\frac{\theta}{2}\right) 
e^{-im\theta}\, \mathrm{Tr}_{\mathcal{H}_-}\left(\mathbb{D}^{-\mu-i\cot\frac{\theta}{2}}\right) \\
&= \frac{1}{4\pi}\int_{-\infty}^{\infty} d\left(\cot\frac{\theta}{2}\right)\, e^{-im\theta}\,   \int_{0}^{\infty} dx\, 
x^{-\mu-i\cot\frac{\theta}{2}} \,\frac{d\psi(x)}{dx}\\
&= \frac{1}{4\pi}\int_{-\infty}^{\infty} d\left(\cot\frac{\theta}{2}\right)\, e^{-im\theta}\,   \int_{1}^{\infty} dx\, 
x^{-\mu-i\cot\frac{\theta}{2}} \, \left(1 - \sum_{\gamma_m} x^{\gamma_m - 1} - \sum_{n=1}^{\infty} x^{-2n-1}\right)
\end{split}
\end{align}
where we have used \cref{sumVM}. As expected this has contribution from the trivial zeroes of the Riemann zeta function 
(the last set of terms) as well as its pole (the first term 1 in the parentheses). We shall discuss the contribution of the trivial 
zeroes shorty (in \cref{ssec:MMGamma}), but for now, in the following let us concentrate on the contributions from the 
\emph{non-trivial zeroes only} (the first sum)\footnote{The divergence from the first term in the integral corresponds to the 
pole at $s=1$. It can be removed by multiplying the partition function by a factor of $(s-1)$. Since our primary goal is to 
look at the non-trivial zeroes, we choose to ignore this issue.}. Splitting these zeroes into real and imaginary parts, 
$\gamma_m = \sigma_m + i t_m$, this piece is
\begin{align}
\begin{split}\label{renBeta3}
\tilde{\beta}_m^{\mathrm{ren}} &=  - \sum_{\gamma_m=\sigma_m + i t_m} \frac{1}{4\pi}\int_{-\infty}^{\infty} 
d\left(\cot\frac{\theta}{2}\right)\, e^{-im\theta}\,   \int_{1}^{\infty} \frac{dx}{x}\, x^{\sigma_m - \mu +
 i(t_m - \cot\frac{\theta}{2})} \\
&=  - \sum_{\gamma_m=\sigma_m + i t_m} \frac{1}{4\pi}\int_{-\infty}^{\infty} 
d\left(\cot\frac{\theta}{2}\right)\, e^{-im\theta}\,   \int_{0}^{\infty} d(\ln x)\, e^{\left(\sigma_m - \mu +
 i(t_m - \cot\frac{\theta}{2})\right)\ln x} 
\end{split}
\end{align}
The renormalised parameters $\beta_m^{\mathrm{ren}}$ should be finite for the matrix model to be defined. The
requirement for this is that the integrals be convergent. The condition for that is met if $\sigma_m \le \mu$ for all 
$\sigma_m$. A priori not any value of $\mu$ is distinguished, however, saturation of the inequality (namely 
$\sigma_m=\mu$ for all $m$) is certainly more elegant. In this case, thanks to the reflection symmetry, the unique 
choice is $\mu = \frac{1}{2}$.

\subsection{Matrix model for gamma function and the trivial zeroes}\label{ssec:MMGamma}
Recall that the adelic zeta function $\zeta_{\mathbb{A}}(s)$ \cref{adelicZ} in \cref{ssec:RevPAdic} (as well as the 
symmetric zeta function $\xi(s)$ \cref{SymmZ}) retain only the non-trivial zeroes. The poles of the gamma function 
(zeta function at infinite place $\zeta_{\mathbb{R}} (s) = \pi^{-s/2} \Gamma\left(\frac{s}{2}\right)$ defined in \cref{zetaRMellin}) 
precisely cancel the trivial zeroes. Let us construct a unitary matrix model which have the information about the poles 
of the gamma function in its eigenvalue density. 

The gamma function $\Gamma(s)$ has simple poles at zero and the negative integers in the $s$-plane. Under the 
conformal map \cref{ConfMap}, it does not have any poles inside the unit circle in the $z$-plane. The resolvent $R(z)$, 
defined in \cref{eq:Rz}, for this model is
\begin{equation}
R(z) = \left\{
\begin{array}{ll}
1 + \displaystyle{\frac{z}{2} \frac{d}{dz}} \ln \left[ \zeta_{\mathbb{R}}\left(\frac{1+z}{1-z}\right) 
\prod_{n=1}^\infty e^{-\frac{1+z}{2n(1-z)}} \right] &\qquad \text{for } |z| < 1\\
\displaystyle{\frac{z}{2} \frac{d}{dz}} \ln \left[ \zeta_{\mathbb{R}}\left(-\frac{1+z}{1-z}\right) 
\prod_{n=1}^\infty e^{\frac{1+z}{2n(1-z)}} \right] &\qquad \text{for } |z| > 1 
\end{array}\right.
\label{GammaRes}
\end{equation} 
{}From the resolvent, one can find the parameters of the UMM 
\begin{align}
\begin{split}
\beta_m^{(\mathbb{R})} &= \frac{1}{2\pi i} \oint \frac{dz}{z^{m+1}}\, R_<(z)\\
&= - \frac{1}{8\pi} \int_{-\infty}^\infty d\left(\cot\frac{\theta}{2}\right)\, e^{-i m\theta}\, 
\left[\frac{\Gamma'(s/2)}{\Gamma(s/2)} - \ln\pi - \sum_{n=1}^\infty\frac{1}{n} \right]_{s=i\cot\frac{\theta}{2}}
\end{split}\label{betaGamma}
\end{align}
where $\theta$ parametrise the unit circle $|z|=1$, as in \cref{ConfMap}. In order to simplify this, let us use the 
Weierstrass product representation
\begin{equation*}
\Gamma\left(\frac{s}{2}\right) = 2 e^{- \frac{1}{2}\gamma s}\, \frac{1}{s} \prod_{n=1}^\infty \left(1 + \frac{s}{2n}\right)^{-1}\, 
e^{\frac{s}{2n}}
\end{equation*}
to write
\begin{equation*}
\frac{d}{ds} \ln \Gamma\left(\frac{s}{2}\right) = - \frac{\gamma}{2} + \sum_{n=1}^\infty \frac{1}{2n} - \int_1^\infty 
dx\, x^{-s} \left( \frac{1}{x} + \sum_{n=1}^\infty x^{-2n-1}\right)
\end{equation*}
where $\gamma = \displaystyle{\lim_{k\rightarrow\infty} \left(\sum_{n=1}^k \frac{1}{k} - \ln k\right)}$ is the Euler-Mascheroni 
constant. Substituting this in \cref{betaGamma}, we find the parameters of the UMM for $\zeta_{\mathbb{R}}$ as
\begin{eqnarray}
\beta_m^{(\mathbb{R})} &=& \frac{1}{4\pi}\!\displaystyle{\int_{-\infty}^\infty}\!\! d\!\left(\!\cot\frac{\theta}{2}\right) 
e^{-i m\theta} \left[ \frac{\ln\pi}{2} + \frac{\gamma}{2} - \frac{i}{\cot\frac{\theta}{2}} + \sum_{n=1}^\infty \left(
\frac{1}{2n + i\cot\frac{\theta}{2}}\right) \right] \label{betaGammaFinal}\\
&=&  \frac{1}{4\pi}\!\displaystyle{\int_{-\infty}^\infty}\!\! d\!\left(\!\cot\frac{\theta}{2}\right) 
e^{-i m\theta} \left[ \frac{\ln\pi}{2} + \frac{\gamma}{2} + \int_1^\infty\! \frac{dx}{x}\, 
x^{-i\cot\frac{\theta}{2}} \left( 1 + \sum_{n=1}^\infty \frac{1}{x^{2n}} \right)\right] \nonumber 
\end{eqnarray}
These coefficients are divergent, however, as in the discussions in \cref{ssec:synth}, this only affects an overall normalisation 
of the partition function, since the $\theta$-dependent parts of the divergence cancel between the $\mathrm{Tr}\, U^n$ and 
$\mathrm{Tr}\, U^{\dagger n}$ terms.

We can also see this if we start with the product of the local zeta $\zeta_p(s)$ for all primes, together with 
$\zeta_{\mathbb{R}}(s)$, the zeta function at infinite place
\begin{equation}
s(s-1)\,\zeta_\infty(s) \prod_p\zeta_p(s) = s(s-1)\, \pi^{-s/2}\Gamma\left(\frac{s}{2}\right) \zeta(s) = 
\prod_m \left(1- \frac{s}{\gamma_m}\right) \label{AllZetas}
\end{equation}
where $\gamma_m$ are the non-trivial zeroes of the Riemann zeta function. Hence, 
\begin{equation*}
\frac{1}{s} + \frac{1}{s-1} + \frac{\zeta^{\prime}_\infty(s)}{\zeta_\infty(s)} + \sum_p 
\frac{\zeta^{\prime}_p(s)}{\zeta_p(s)} = \sum_m \frac{1}{s-\gamma_m}
\end{equation*}
Using \cref{betaGamma} and the results in \cref{ssec:Rzprop}, together with the identity 
$\displaystyle{\frac{1}{s-s_0}} = \displaystyle{\int_1^\infty x^{s-s_0-1}}$ (valid for $\mathrm{Re}(s-s_0)>0$), 
one finds
\begin{eqnarray*}
&{}& \beta_m^{\mathrm{eff}} \:=\: \beta_m^{(\mathbb{R})} + \sum_p \ln p\, \beta_m^{(p)} \\ 
&=& \!\!\int_{-\infty}^\infty d\bigg(\!\cot\frac{\theta}{2}\bigg)\frac{e^{-im\theta}}{4\pi}  \left[ \ln\sqrt\pi + \frac{\gamma}{2} + 
\int_1^\infty\!\! dx\, x^{-i\cot\frac{\theta}{2}} \left(1 + \frac{1}{x} - \sum_{n} x^{\gamma_n-1} + \sum_p\ln \sqrt{p}\right) \right]
\end{eqnarray*} 
which is essentially the coefficients found in \cref{ssec:Renorm} corresponding to the non-trivial zeroes, but without 
the renormalisation. In fact, this is what one would expect from \cref{AllZetas}. 

\subsection{Towards a Hamiltonian in the large phase space}\label{ssec:lpsHamiltonian} 
We saw in \cref{ssec:LocHamiltonian} that (the fluctuating part of) the phase space density of the unitary matrix model 
for the local zeta function at a prime $p$ suggests an $xp$-type Hamiltonian $H_{(p)} = \cot\left(\frac{\theta}{2}\right)
\ln D_{(p)}$. For the matrix model corresponding to the Riemann zeta function, it is natural to propose the Hamiltonian 
\begin{equation}\label{schematicH}
H = \cot\left(\frac{\theta}{2}\right)\ln\mathbb{D} \sim \cot\left(\frac{\theta}{2}\right)\sum_p \ln D_{(p)}
\end{equation}
where the definition of $\ln \mathbb{D}$ is similar to that is \cref{bigDalpha}. 
This is only schematic, as it may lead to divergences. Indeed, we have seen that the parameters $\beta_m$ of the 
combined matrix model are divergent, and needs a renormalisation before the can make sense. It would not be a 
surprise if after combining the data for the local models, the full Hamiltonian only makes sense as a 
\emph{quantum operator}.

Let us start with the quantum Hamiltonian $\hat{H}=\hat{x}\hat{p}$, where $\hat{x}$ and $\hat{p}$ satisfy the 
canonical commutation relation $\left[\hat{x},\hat{p}\right]=i$ (in units $\hbar=1$). One may define another 
Hamiltonian $\hat{H}'$ by a similarity transform
\begin{equation}
\hat{H}' = e^{\mu\hat{p}}\,\hat{x}\hat{p}\,e^{-\mu\hat{p}} = \left(\hat{x}-i\mu\right)\hat{p}
\label{simTrH}
\end{equation}
Both $\hat{H}$ and $\hat{H}'$ lead to the same classical Hamiltonian. Thus, for $\mu\in\mathbb{R}$, one gets 
a class of Hamiltonians that differ by a complex shift in $\hat{x}$. The action of this similarity transformation on 
the operator $e^{-i\hat{H}}$ is as follows
\begin{equation}
e^{\mu\hat{p}}\,e^{-i\hat{x}\hat{p}}\, e^{-\mu\hat{p}} = \exp\left(-i\hat{x}\hat{p} - \mu\hat{p}\right)
\label{simTrRho}
\end{equation}
We shall now propose a construction for the operators on the Hilbert space spanned by the Kozyrev wavelets. 
{}From its spectrum, we see that the momentum operator is the logarithm of the Vladimirov derivative, but we need 
to identify the generalised coordinate. To this end, let us note that the recursive construction of the wavelet basis 
defines the raising and lowering operators $J^{(p)}_\pm$. Together with (the log base $p$ of) the Vladimirov derivative, 
one finds the following SL(2,$\mathbb{R}$) algebra and its action on the wavelets
\begin{align}\label{SL2Symmetry}
\begin{split}
J^{(p)}_\pm \left| \psi_{-n+1,m,j}\right\rangle = \pm n \left| \psi_{-(n\pm 1)+1,m,j}\right\rangle, \quad
&\quad \log_p D_{(p)} \left| \psi_{-n+1,m,j}\right\rangle = n \left| \psi_{-n+1,m,j}\right\rangle  \\
\left[ J^{(p)}_+, J^{(p)}_-\right] = 2\log_p D_{(p)},\quad &\quad \left[\log_p D_{(p)},J^{(p)}_\pm\right] = \mp J^{(p)}_\pm
\end{split}
\end{align}
as shown in \cite{DuGhLa}. Consider a Schwinger realisation of this algebra using a pair of (bosonic) creation and 
annihilation operators $a_{(p)}^\dagger, a_{(p)}$ and $b_{(p)}^\dagger, b_{(p)}$ (satisfying $[a_{(p)}, a_{(p)}^\dagger] 
= 1$ and $[b_{(p)}, b_{(p)}^\dagger] = 1$) in terms of which
\begin{equation}\label{Schwinger}
J^{(p)}_+ = a_{(p)}^\dagger b_{(p)},\quad J^{(p)}_- = b_{(p)}^\dagger a_{(p)}\quad\text{and}\quad \log_p D_{(p)} = 
\frac{1}{2}\left(b_{(p)}^\dagger b_{(p)} - a_{(p)}^\dagger a_{(p)} \right)
\end{equation}
Since the momentum operator $\hat{P}_{(p)}$ is the difference of the two number operators $N$, it is natural 
to identify the coordinate $\hat{X}_{(p)}$ with the difference in the phase operator $\Phi$ (such that $\left[\Phi,
N\right] = i$)
\begin{equation}
\hat{P}_{(p)} = \ln D_{(p)}  = \frac{\ln p}{2}\left(N_b - N_a \right)\quad
\text{and}\quad \hat{X}_{(p)} = \frac{1}{\ln p}\left(\Phi_b - \Phi_a\right) 
\end{equation}
The construction of the phase operator has a long history, and is somewhat indirect (see \cite{MaRhodes:2015} 
and references therein, where the construction of analytic functions of $\Phi$ has also been discussed). The coordinate 
assumes the continuous values $\cot\frac{\theta}{2}$, and the momentum has a discrete spectrum. The Hilbert space 
that is relevant for us, is spanned by only the positive values of $\ln D_{(p)}$, therefore, it should suffice to `freeze' one 
of the oscillators.

\subsection{Wigner function in the phase space}\label{ssec:Lps}
What is the phase space? Although a proper understanding of its geometry and analytical structure will need further 
investigation, let us take a first step in this direction. The Hilbert space $\mathcal{H}^{(p)}_-$ (in the Koopman-von 
Neumann formalism) for the matrix model at a prime $p$, consists of a suitably truncated subspace of complex 
valued, mean-zero, square integrable functions on a compact subspace of $\Qp$. The large Hilbert space 
$\mathcal{H}_- = \bigotimes_p \mathcal{H}^{(p)}_-$ is a consistent truncation of functions with these properties on 
a compact subspace of the product space $\bigotimes_p\Qp$.

For a local matrix model, the coordinate and momentum take values in the set $\mathbb{R}\times\mathbb{Z}$, where 
the first factor is from the eigenvalues and the second corresponds to the number of boxes in a Young diagram in a 
representation. In expressing (the fluctuating part of) the phase space density as the trace of an operator, however, 
the topology in the momentum space was determined by the ultrametric norm of the $p$-adic numbers. Thus, the local 
phase space should be taken as $\mathbb{R}\times\Zp$. When we combine these local constituents to define the large 
phase space, the Hamiltonian is the sum of the prime factors acting on vectors in the large Hilbert space. Hence the 
proposed large phase space is
\begin{equation}\label{lpsGeom}
\bigotimes_{p\in\mathrm{primes}} \big(\mathbb{R}\times \Zp \big) \sim \mathbb{R}\times \left(\otimes_p \Zp \right)
\end{equation}
where in the second expression, we have taken the coordinate as the `diagonal' in $\otimes_p\mathbb{R}$, as it is 
common for all $p$. This space is reminiscent of the adelic space. 

In order to understand this large phase space better, let us go back to the small phase space for which our description 
has been classical. We shall try to understand the corresponding quantum system. {}From the spectrum of the momentum, 
we can label the eigenstates of the momentum operator $\hat{P}_{(p)}=\ln D^{(p)}$ (see \cref{ssec:LocHamiltonian}) as 
$|n_{(p)}\rangle \equiv |\psi^{(p)}_{-n+1,0,1}\rangle$ with eigenvalues\footnote{In what follows, we shall sometime omit the
label $p$ to avoid cluttering up formulas.} $n\ln p$ for $n\in\mathbb{N}$ because we are restricted to the subspace 
$\mathcal{H}^{(p)}_-$. The eigenstates of the position operator $\hat{X}_{(p)}$, denoted by $|x_{(p)}\rangle$ (restricted to 
$\mathcal{H}^{(p)}_-$ by the projection operator $\mathcal{P}_- = \sum_{n\in\mathbb{N}} |n\rangle\langle n|$) may be 
expanded as  
\begin{equation}
\left|x_{(p)}\right\rangle = \sum_{n_{(p)}\in\mathbb{N}} e^{i x_{(p)} n_{(p)}\ln p} \left|n_{(p)}\right\rangle
\label{Xpansion}
\end{equation}
It is worth emphasising that the sum is only over positive integers, and the same is true of the inner product
\begin{equation}\label{pPsiX}
\left\langle x'_{(p)}|x_{(p)}\right\rangle = \sum_{n\in\mathbb{N}} e^{i (x_{(p)} - x'_{(p)}) n\ln p} 
\end{equation}
Using the identity \cref{IdentityInfSum}, this can alternatively be written as   
\begin{equation}\label{pPsiX2}
\left\langle x'_{(p)}|x_{(p)}\right\rangle = \sum_{m\in\mathbb{Z}} \int_0^\infty d\kappa\, 
e^{i\kappa\left(x_{(p)} - x'_{(p)} - \frac{2\pi m}{\ln p}\right)} = \Psi_{x_{(p)}}^{(p)}(x'_{(p)}) 
\end{equation}
Either of the forms is consistent with the periodic nature of (the difference of) the phase operator. Equivalently, with the 
fact that the eigenvalues of the UMM, which are the poles of the local zeta function, repeat periodically along the imaginary 
axis. We also note that the above is the Schr\"{o}dinger wavefunction $\Psi_x^{(p)}(x')$ in the position representation. 

Let us construct the Wigner distribution function $W^{(p)}_{x'}(x,q)$ for the position eigenstate labelled by $x'$, which is 
defined in terms of the operator $\hat{\rho}^{(p)}_{x'} = |x'\rangle\langle x'|$ as
\begin{align}
\begin{split}
W_{x'}^{(p)}(x,q) &= \displaystyle\int_{-\infty}^{\infty} dy \left\langle x +\frac{y}{2} \right| \hat{\rho}^{(p)}_{x'}
\left| x - \frac{y}{2}\right\rangle\, e^{iyq}\\
&= -2\int_0^\infty d\kappa\, e^{2i(q-\kappa)(x-x')}\, \frac{d\jmath_p(e^{\kappa})}{d\kappa}  
\frac{d\jmath_p(e^{2q-\kappa})}{d\kappa}  
\end{split}\label{pWigner}
\end{align}
where we have used the wavefunction \cref{pPsiX}, performed the integrals over $q$ and one of the $\kappa$'s, and used 
the identity \cref{IdentityInfSum} and the definition of the local counting function $\jmath_p(x)$. The above form requires 
that $q > \kappa/2$, and since $0 \le \kappa < \infty$, this is the expression in $q \ge 0$. Before we proceed further it would 
be prudent to verify that the same Wigner function is obtained starting with its momentum space definition 
\begin{equation}
W^{(p)}_{x'}(x,q) = \int_{-\infty}^\infty d{q'} \left\langle q + \frac{q'}{2} \bigg|\, x'\right\rangle \left\langle x'\, \bigg| 
q - \frac{q'}{2}\right\rangle\, e^{-i x q'} \label{pWignerMom}
\end{equation}
{}From \cref{Xpansion} we see that, upto a factor of $\ln p$, scalar products between the position and momentum 
eigenstates are supported only at positive integer points on the real line, that is
\begin{equation*}
\left\langle q + \frac{q'}{2} \bigg|\, x'\right\rangle = e^{ix' \left(q+\frac{q'}{2}\right)}\, \sum_{n\in\mathbb{N}}
\delta\left(q+\frac{q'}{2} - n\ln p\right)
\end{equation*}  
It is easy to check that this will indeed lead to the same Wigner function as in \cref{pWigner} after we substitute 
the scalar product above and redefine of the dummy variable of integration in the expression \cref{pWignerMom}.
  
The integral of the Wigner function over momentum should give the probability density in the coordinate space. Let us 
check that this is the case\footnote{The divergent constant below is a result of working with unnormalized wavefunction. 
The issue is similar to the normalisation of plane waves, where one uses a delta-function normalisation or puts the 
system in a box.}.
\begin{align}
\begin{split}
\int dq\, W_{x'}^{(p)}(x,q) &= \sum_{n\in\mathbb{N}} \sum_{m\in\mathbb{N}} e^{i(x - x')(m-n)\ln p}\\
& = \frac{2\pi}{\ln p} \left(\sum_{n\in\mathbb{N}} 1\right) \sum_{m\in\mathbb{Z}} \delta\left(x - x' - \frac{2\pi m}{\ln p}\right)
\end{split}
\label{prhoWigner}
\end{align}
We see the periodicity of the eigenvalue distribution, related to the poles of the local zeta function. The sum of the 
exponentials relates to the density obtained from the resolvent, however, let us compare with what we got from the 
eigenvalue distribution of the matrix model. To see the equivalence, let us start from \cref{pZetaRZed} with 
\cref{ConfMap}, and do an inverse Mellin transformation on \cref{locJZeta} to get
\begin{equation*}
\frac{d\jmath_p(x)}{dx} = \frac{1}{2\pi i} \int_{a-i\infty}^{a+i\infty} \frac{p^{-s}x^{s-1}}{1-p^{-s}}\,ds = 
\frac{1}{\ln p} \sum_{n\in\mathbb{Z}} x^{\frac{2\pi i n}{\ln p}-1}
\end{equation*}
Hence
\begin{align*}
\begin{split}
\int_0^\infty x^{-i\cot\frac{\theta}{2}} \frac{d\jmath_p(x)}{dx} \, dx &= \frac{\pi}{\ln p} \sum_{n\in\mathbb{Z}} 
\delta\left(\cot \frac{\theta}{2} - \frac{2\pi  n}{\ln p}\right)\\
&\qquad - \frac{i}{2\ln p}\sum_{n\in\mathbb{Z}} \frac{ 2 \, - \displaystyle\lim_{\Lambda\rightarrow\infty} 
\left( e^{i\Lambda \left(\cot \frac{\theta}{2} - \frac{2\pi  n}{\ln p}\right)} - e^{-i\Lambda\left( \cot\frac{\theta}{2} - 
\frac{2\pi  n}{\ln p} \right)} \right)}{\cot \frac{\theta}{2} - \frac{2\pi n}{\ln p}}\\
&= \frac{\pi}{\ln p} \sum_{n\in\mathbb{Z}} \delta\left( \cot \frac{\theta}{2} - \frac{2\pi  n}{\ln p}\right) - 
\frac{i}{\ln p\,\cot \frac{\theta}{2}} - \frac{i}{\ln p} \sum_{n=1}^\infty \frac{2\cot \frac{\theta}{2}}{\cot^2
\frac{\theta}{2} - \left(\frac{2\pi n}{\ln p}\right)^2} 
\end{split}
\end{align*}
In the above, we have omitted the terms involving the limit $\Lambda\to\infty$, as the contributions of these terms
to an integral with any well-behaved function will vanish. Therefore, the fluctuating part of the eigenvalue density is
\begin{equation}\label{mmside}
\rho_{\text{fl}}(\theta) = \frac{\pi}{\ln p} \sum_{n\in\mathbb{Z}} \delta\left(\cot \frac{\theta}{2} - \frac{2\pi  n}{\ln p}\right)
\end{equation}
We see that with $x-x'=\cot\frac{\theta}{2}$, \cref{prhoWigner} matches exactly with the above, provided we identify
the divergent factors. 

On the other hand, after integrating the Wigner function over position, we find
\begin{equation}
\int_{-\infty}^\infty dx\, W^{(p)}_{x'}(x,q) = 2\left( \frac{d\jmath_p(e^{q})}{dq} \right)^2 = 2\delta(0)\, 
\sum_{n\in\mathbb{N}} \delta\left(q - n\ln p\right)
\label{XIntWinger}
\end{equation}
which is related to the momentum distribution (upto a divergent factor). The divergence seems to suggest that the 
Wigner function ought to be `renormalised', which could perhaps be achieved by a similarity transform.

\bigskip

In the large phase space, the position eigenstates in $\mathcal{H}_-$, when expanded in the momentum basis, after
the conventional scaling as in \cref{lHs}, is
\begin{equation*}
\left| x \right\rangle = \sum_{p\in\text{primes}}\sqrt{\ln p}\, \sum_{n_{(p)}\in\mathbb{N}} e^{i x n_{(p)}\ln p} \left| n_{(p)}
\right\rangle
\end{equation*}
where $\left\langle n'_{(p')} | n_{(p)} \right\rangle = \delta_{n_{(p)} n'_{(p')}}\delta_{pp'}$ is the normalization of the 
momentum states. The inner product of these states that follows from the above is
\begin{equation}\label{PsiX}
\left\langle x' | x \right\rangle = \sum_p \ln p\, \sum_{n_{(p)}\in\mathbb{N}} e^{i (x-x') n_{(p)}\ln p}
\end{equation}
The above can be re-expressed in two different ways. Either we may introduce integrals for each $p$, as in 
\cref{pPsiX2}, or we may use only one integral over $\kappa$, leading to the two following expressions which 
ought to be equivalent.
\begin{equation}\label{PsiXalt}
\left\langle x' | x \right\rangle = \sum_p \ln p \sum_{m_{(p)}\in\mathbb{Z}} \int_0^\infty d\kappa_{(p)}  
e^{i \left(x-x' + \frac{2\pi m_{(p)}}{\ln p}\right)\kappa_{(p)}} = \int_0^\infty d\kappa\, e^{i(x - x')\kappa}\, 
\frac{d\psi(e^\kappa)}{d\kappa}
\end{equation}
In the above, $\psi(x)$ is the summatory von Mangoldt function \cref{sumVM} that is related to the counting of prime 
numbers. We also note that since $x$ is chosen to be the same for all primes, that is along the diagonal in $\otimes_p
\mathbb{R}$, one would not expect to have any periodicity since the periods along the different directions are 
incommensurate. 

Hence we see that when we combine the information from all primes to get the Wigner function in the large phase space,
there are two ways of parametrising it. One way is to think of the coordinates and momenta as infinite component 
vectors $(\mathbf{x},\mathbf{q}) = \left((x,x,\cdots),(q^{(2)},q^{(3)},\cdots)\right)$. The other is to simply take $(x,q) \in
\mathbb{R}^2$. The coordinates for all primes have been identified, thus justifying the variable $x$. For the canonically 
conjugate momentum operator $P=\sum_p\ln D^{(p)}$ on the other hand, its eigenvalues (and hence expectation values)
are real numbers. Thus, by the correspondence principle, the momentum may be taken to be real valued. In the following, 
we shall work with the latter choice, that is, take the classical phase space of the matrix model for the Riemann zeta 
function to be $\mathbb{R}^2$. There may be a deeper connection between the two descriptions. In this context, it may 
be relevant to mention that Ref.\cite{Neretin2006} shows a canonical bijection between the spaces of locally constant 
complex valued functions on $\otimes_p\Qp^n$ and distributions on $\mathbb{R}^n$. It will be worth exploring if this 
result (for $n=1$) explains the equivalence of the two ways of parametrising the large phase space.

Let us start with the Wigner function on the large phase space for the operator $\hat{\rho}_{x'+i\mu} = \sum_p \ln p \,
\hat{\rho}_{x'+i\mu}^{(p)}$, in which we have included the conventional additional factor of $\ln p$ to be consistent with 
\cref{BigBetaM}, and at the same time have shifted $x'\rightarrow x' + i\mu$ in anticipation.
\begin{align}
\begin{split}
W_{x'+i\mu} (x,q) &= \displaystyle\int_{-\infty}^{\infty} dy\, \left\langle x +\frac{y}{2} \Big| x + i\mu\right\rangle \left\langle 
x  - i\mu \Big| x - \frac{y}{2} \right\rangle\, e^{iy q}\\
&= -2 e^{-2\mu q - 2iq (x-x')}\, \int_0^\infty d\kappa\, e^{-2i\kappa(x - x')}\, \frac{d\psi(e^\kappa)}{d\kappa}\,
\frac{d\psi(e^{2q-\kappa})}{d\kappa}  
\end{split}\label{lpsWigner}
\end{align}
When integrated over $x$, we get
\begin{equation}
\int_{-\infty}^\infty dx\, W_{x'+i\mu} (x,q) = \left(e^{-\mu q}\, \frac{d\psi({e^q})}{dq}\right)^2 
\label{lpsWigXint} 
\end{equation}
which involves the square of the derivative of $\psi(x)$. Since the integrated Wigner function is the probability density, 
the conclusion to draw from here is that $e^{-\mu q}\frac{d\psi^(e^q)}{dq}$ is the wavefunction (in momentum space).   
On the other hand, when integrating over $q$, if we take 
\begin{equation}
\frac{d\psi(e^q)}{dq} = \sum_p \ln p \sum_{n\in\mathbb{N}}\delta(q - n \ln p) \label{psiDer}
\end{equation}
we get
\begin{align}
\begin{split}
\int_{-\infty}^\infty dq\, & W_{x'+i\mu} (x,q) = \sum_p \ln p \sum_{p'} \ln p' \sum_{n\in\mathbb{N}} \sum_{m\in\mathbb{N}}
e^{-\mu(n\ln p + m\ln p')} e^{-i(x - x')(n\ln p - m\ln p')}\\
&= 2\pi \sum_p \ln p \left(\sum_{n\in\mathbb{N}} e^{-\mu n\ln p}\right) \sum_{m\in\mathbb{Z}}
\delta \left(x - x' - \frac{2\pi m}{\ln p}\right) + \sum_{p\ne p'}\left(\cdots\right)
\end{split}\label{lpsWigQint} 
\end{align}
{}From what we have argued, this should give the eigenvalue density of the matrix model of the Riemann zeta function,
and is in fact the square of wavefunction (in position space). However, after combining all primes, this distribution does 
not have any periodicity, unlike each of its prime factors. Recall that we have identified the coordinates for the each of the 
prime factors, therefore, the coordinate in the large phase space is along the `diagonal' $\mathbb{R}$ in $\otimes_p 
\mathbb{R}$. Hence one misses the periodically placed poles of the local zeta factors along this direction. The second 
terms in the sum involving the cross terms should vanish, since sectors labelled by different primes are `orthogonal'. If 
we assume the Riemann hypothesis, this function should have $\delta$-function support on the non-trivial zeroes of the 
Riemann zeta. We shall now try to argue that this indeed is the case.
 
\subsection{Trace formula and the Wigner function}\label{ssec:TrForm}
In order to understand the distribution, it is useful to study its effect on a smooth function. With this objective, for a smooth 
function $g(q)$ let us consider the integral 
\begin{equation*}
\int_{-\infty}^\infty g(q) e^{-q/2}\,\frac{d\psi(e^q)}{dq}\, dq = \sum_p \ln p \sum_{n\in\mathbb{N}} p^{-n/2}\,
g(n\ln p)
\end{equation*}
where we have used \cref{psiDer} to get the RHS. In the LHS, first we change the argument of $\psi$ to $q'$ by introducing 
a $\delta(q-q')=\int du\, e^{iu(q-q')}$, and then use the expansion \cref{sumVM} in terms of Riemann zeroes to get
\begin{equation*}
\int_{-\infty}^\infty du\, h(u) \int_0^\infty dq' \left( e^{\left(\frac{1}{2} - iu\right)q'} + e^{-\left(\frac{1}{2} + iu\right)q'} - 
\sum_{\gamma_m} e^{\left(\gamma_m - \frac{1}{2} - iu\right)q'} - \sum_{n=0}^\infty e^{-\left(2n + \frac{1}{2} + iu\right)q'}
\right) 
\end{equation*}
where $h(u) = \int g(q) e^{iqu} dq$ is the Fourier transform of $g(q)$ and the lower limit on the $q'$ integral is $0$ because 
$\psi$ vanishes for negative arguments. Now changing the signs $u\to -u$ and $q'\to -q'$ in the first three terms of the
integrand and using the fact that the zeroes $\gamma_m$ are symmetrically distributed, we can combine the two 
expressions of the LHS to extend the limit of the $q'$ integral over the real line. For the last term in the integrand that
involves the sum over integers, we use manipulations similar to \cref{ssec:MMGamma}. After performing the integrals,
we finally arrive at
\begin{align}
\begin{split}
h\bigg(\frac{i}{2}\bigg) + h\bigg(\!\!-\frac{i}{2}\bigg) - & \sum_{t_m} h\left(t_m \right)
+ \int_{-\infty}^\infty\!\!\!\! du\,h(u) \frac{\Gamma'\left(\frac{1}{4}+i\frac{u}{2}\right)}{\Gamma\left(\frac{1}{4} +
i\frac{u}{2}\right)} \\
&= g(0) \ln\pi +  2\sum_p \ln p \sum_{n\in\mathbb{N}} p^{-\frac{n}{2}} g(n\ln p) 
\end{split} \label{TrForm} 
\end{align}
where we have used $\gamma_m=\frac{1}{2}+it_m$. This is an example of what are known as trace formulas 
\cite{hejhal1976}. 

We see from its definition \cref{lpsWigner} that $\int dq\, W_{x'}(x,q) = \left|\Psi_{x'}(x)\right|^2$ in the square of the
wavefunction \cref{PsiXalt} (with $x'$ shifted by $i\mu$). Hence from the trace formula we find that the wavefunction is
\begin{eqnarray*}
\Psi_{x'+i\mu}(x) &=& \frac{1}{2}\left[ \delta\bigg(x-x'-\frac{i}{2}\bigg) + \delta\bigg(x-x'+\frac{i}{2}\bigg)  \right. \\
&{}& \qquad\qquad \left. - \sum_{t_m} \delta(x-x'-t_m) + \frac{\Gamma'\left(\frac{1}{4} + 
\frac{i}{2}(x-x')\right)}{\Gamma\left(\frac{1}{4} + \frac{i}{2}(x-x')\right)} - \ln \pi\right]  
\end{eqnarray*}
provided $\mu=\frac{1}{2}$. In the above, a Dirac delta function with a complex argument is to be understood in terms of 
the residue of the function that appears in the integrand with it. One would expect the terms involving the $\Gamma$-function 
to be accounted for by the contribution from $\zeta_{\mathbb{R}}$. Although we have not verified it, we think that it would be 
natural for the wavefunction for the adelic zeta function \cref{adelicZ} to involve only $\sum_{t_m} \delta(x-x'-t_m)$ that
correspond to the non-trivial zeroes. Working backwards, one can see that this would be equivalent to the following 
inner product
\begin{equation}\label{IPLargePS}
\left\langle x' + i\mu | x \right\rangle = \sum_{m\in\mathbb{Z}} \delta\left(x - x' - i(\mu - \gamma_m)\right)
= \sum_{m\in\mathbb{Z}} \delta\left(x - x' - t_m\right) 
\end{equation}
where, the last expression is true for $\mu = \frac{1}{2}$. This may be realised through the following similarity transform 
(defined with a star-product appropriate to the phase space formalism) 
\begin{equation}
W_{x' + i\mu} (x,q) = e^{\mu q} \star W_{x'} (x,q) \star e^{-\mu q}  
\end{equation}
of the naive Wigner function one would have assumed from the individual prime factors (together with the contributions from
the Gamma function). This is consistent with the similarity transform \cref{simTrH} on the Hamiltonian, and the fact that the 
support of the eigenvalues of the UMM for the Riemann zeta function is shifted from those of the local zeta functions. The 
square of the wavefunction \cref{IPLargePS} is again of the same form, upto an infinite factor, as in the case of the Wigner 
function for a fixed $p$.  
 
\section{Conclusion, summary and outlook}\label{sec:concl}
We have initiated a programme to arrive at a Hamiltonian corresponding to the non-trivial zeroes of the Riemann zeta 
function. That this is a possibility was suggested by Hilbert and P\'{o}lya. A concrete proposal was made by Berry and 
Keating\cite{BK,BerryKeating2} (see also \cite{Sierra:2007du,ConnesEssay,ConnesTrForm}), and been refined by many 
others since. Very recently a similarity transform of the Berry-Keating Hamiltonian was proposed in \cite{Bender:2016wob},
where, the authors show that upon enforcing specific boundary conditions the eigenvalues of the Hamiltonian coincide with 
the zeroes of the Riemann zeta function. The corresponding eigenfunctions are the Hurwitz zeta functions. Their approach 
does not have any explicit connection with prime factorisation or the adelic structure. We also refer to \cite{Bellissard:2017}  
for a critique and \cite{Bender:2017} for a reply to it. (Let us also mention Ref.\cite{Schumayer:2011yp}, and references 
therein, for physicists approaches to various aspects of the Riemann zeta function.) 
We approach this problem using the phase space description of the unitary matrix model (UMM), a random ensemble of 
unitary matrices, the eigenvalue distribution of which is known to exhibit the same statistical features as the Riemann 
zeroes \cite{montgomery,sarnak1,Odlyzko}. 

In fact a phase space description for a UMM for the Riemann zeta function was developed in \cite{Dutta2016byx}, however, 
it was not possible to arrive at a Hamiltonian by working with $\zeta(s)$. In this paper, we have followed the same general 
idea, but split the analysis into two parts. First we have considered the `prime factors' of the zeta function, and constructed 
UMMs for each prime factor. That is, a UMM for each of the local zeta function at a prime $p$. Each of these zeta functions 
has regularly spaced poles on the imaginary axis. We have then combined these UMMs to define a UMM for the Riemann 
zeta function. We encounter divergences in this process. In particular, the coefficients that define the UMM involve an 
infinite sum, and diverge in this process. We propose a `renormalisation' to extract a finite answer to properly define the 
UMM. The coefficients that we get through this process match with those found in Ref.\cite{Dutta2016byx}, which 
constructed a UMM working directly with the symmetric zeta function. 

As in the Berry-Keating proposal, the Hamiltonian we arrive at has the $xp$ form, however, a direct comparison is 
difficult for many reasons. First, we work in the Koopman-von Neumann formalism in which a classical system has 
a Hilbert space description. Secondly, our approach involves the local fields $\Qp$ of $p$-adic numbers and complex
valued functions over these spaces, since we work with the local zeta functions. Thirdly, in trying to make sense of 
the divergent answer, we have to do a similarity transform on this basic Hamiltonian. These features could be 
advantageous, since the $xp$ Hamiltonian cannot be right in its original form. In fact, the truncation to a subspace
of the Hilbert space of complex valued square integrable functions on $\Qp$ that we have is reminiscent of the truncation 
of the allowed region of the classical phase space in Ref.\cite{BK}. (We also recall the work of Connes \cite{ConnesEssay}
in which $p$-adic numbers and the adelic ring play essential roles.) 

It is also important to stress that in this approach, the operator corresponding to the non-trivial zeroes of the 
Riemann zeta function (or the poles of the local zeta functions) is, by construction, the position operator, and not the 
Hamiltonian. However, it is the Hamiltonian that has information about the density of states in the phase space, has the 
Berry-Keating form.

We have only initiated a programme here, but certainly do not claim to have resolved all the issues satisfactorily. There are 
several points that will need further clarity and resolution. First among these is the density function for the eigenvalues, which 
exhibits an essential singularity at $z=1$ (equivalently $\theta=0$). Formally, this makes the eigenvalue density non-positive. 
The issue has its root in the fact that there are an infinite number of poles of the local zeta function (respectively, the zeroes 
of the adelic or symmetric zeta function) beyond any finite value of $\mathrm{Im}(s)$ as it approaches $\pm i\infty$, with 
$\mathrm{Re}(s)$ fixed at $0$ or $1/2$, as appropriate. After the conformal mapping of the critical line in the $s$-plane to 
the unit circle, needed for description in terms of unitary matrices, this results in an accumulation point in the finite part of 
the $z$-plane. This problem would not arise if one could develop a phase space description for the partition function of a 
random ensemble of Hermitian matrices, however, presently the phase space picture (in terms of eigenvalues and 
representations) is available only for a unitary ensemble. We discuss a way to extract a positive density through a regulator 
in \cref{app:NormResolvent}. Secondly, only the fluctuating part of the corresponding phase space density can be related to 
the Hamiltonian. This, however, is not different from the situation in Ref.\cite{BK}. Thirdly, the geometrical and other properties 
of the phase space need a thorough investigation, so that the nature of the Hamiltonian and its spectrum could be understood 
better and directly, independent of the matrix model. Related to this are the issues of self-adjointness or otherwise of these 
operators. It would be desirable to analyse the quantum mechanical problem. This is possible in principle since we have a 
concrete proposal with an explicit Hamiltonian and the Hilbert space on which it acts as an operator. However, the existing 
approaches to quantum mechanics on local fields (see e.g., \cite{Ruelle:1989jv,Dragovich:1995qr}) are not directly applicable, 
since in our proposal the coordinates are real valued while the momenta take values in $\Zp$ or its adelic counterpart, or, to 
be precise the topology (distances and measures) in this part is determined by these ultrametric spaces. This would fit with 
a symplectic structure, if one exists, in the adelic space. Finally, the local zeta functions at primes $p$ (as well as that at the 
`infinite place', i.e., $\mathbb{R}$) have an infinite sequence of poles repeating with regular periodicities. After combining to 
form the large matrix model, these poles disappear, and the special points are the zeroes of the Riemann zeta function. It 
would be instructive to understand how this transformation takes place.   

We close by emphasising the main ingredients of our approach once again. First, we construct a one-plaquette
unitary matrix model given a distribution of special points (poles or zeroes of a function), and make use of its phase
space formulation. We apply this to the prime factors in the Euler product representation of the Riemann zeta 
function, and express the phase space density as the trace of an operator on a suitable Hilbert space of functions
on a subspace of the $p$-adic number field. This operator leads to the suggested Hamiltonian for the zeta function.
It is important to remember that this is not an attempt to prove the Riemann hypothesis, but to find a Hamiltonian
related to its zeroes. 

 
\bigskip

\noindent{\bf Acknowledgments:} PD and DG were supported in part by the research grant no.~5304-2, 
{\em Symmetries and Dynamics: Worldsheet and Spacetime}, from the Indo-French Centre for Promotion 
of Advanced Research (IFCPAR/CEFIPRA), and that of SD by the grant no.~EMR/2016/006294 from the 
Department of Science \&\ Technology, Government of India. DG is thankful for the hospitality at the \emph{VI-th 
International Conference on $p$-adic Mathematical Physics} held at Mexico City, where some preliminary 
results were presented. We thank the participants, especially V.~Anashin, A.~Bendikov, S.~Jeon, V.~Osipov, 
and W.~Zu\~{n}iga-Galindo, for many useful comments. We would also like to thank V.P.~Gupta, C.~Imbimbo, 
A.~Lala, R.~Ramaswamy and R.~Shah. AC, SD and DG acknowledge the hospitality during the workshop 
\emph{Nonperturbative and numerical methods in quantum gravity, string theory and holography} (Code: 
ICTS/NUMSTRINGS/2018/01), at the International Centre for Theoretical Sciences, Bengaluru, where the 
a preliminary draft of this article was prepared. SD and DG acknowledge, respectively, a Simons Associateship
of the Abdus Salam ICTP, Trieste, Italy, and the Albert Einstein Institute, Potsdam, Germany for hospitality during 
the final stages of the work. We would like to thank the anonymous referees for their detailed critique that lead to 
a much improved version.
 

\bigskip

\appendix
\section{Comparison with the matrix model of Ref.\cite{Dutta2016byx}}\label{app:CompPS}  
A one-plaquette unitary matrix model for the symmetric zeta function \cref{SymmZ} was constructed in Ref.
\cite{Dutta2016byx}. The parameters $\beta_m$ were found to be given by the Li coefficients \cref{LiNumbers}. Let us 
compare these with our renormalised $\beta_m^{\mathrm{ren}}$ coefficients obtained in \cref{ssec:Renorm}. Notice 
that for obvious reasons the conformal map \cref{ConfMap} from the $z$- to the $s$-plane that we have used is different 
from the one in \cite{Dutta2016byx}. In order to facilitate comparison one may start by a translation so that the poles of the 
local zeta function lie on the line $\mathrm{Re }(s)=s_0$ and then map the shifted line to the unit circle. This changes the 
expression of the coefficients in \cref{renBeta2} to
\begin{eqnarray}
\beta_m^{\mathrm{ren}}&=& \frac{1}{4\pi}\int_{-\infty}^{\infty} d\!\left(\cot\frac{\theta}{2}\right)\, e^{-im\theta}\,   
\int_{0}^{\infty} dx\, x^{-s_0 - \frac{i}{2}\cot\frac{\theta}{2}} \, \left(1 - \sum_{\gamma_m} x^{\gamma_m - 1} - 
\sum_{n=1}^{\infty} x^{-2n-1}\right) \nonumber\\
&=& - \int_{0}^{2\pi} \frac{d\theta}{2\pi} \frac{e^{-i(m-1)\theta}}{(1-e^{i\theta})^2} \int_{0}^{\infty} 
\frac{dx}{x^{s_0 + \frac{i}{2}\cot\frac{\theta}{2}}} \left( - \int_{a-i\infty}^{a+i\infty} \frac{ds}{2\pi i} x^{s-1} \frac{d}{ds}
\ln \zeta(s)\right) \nonumber\\
&=&\frac{1}{2\pi i}\int_{-\pi}^{\pi} \frac{d\theta}{2\pi} \frac{e^{-i(m-1)\theta}}{(1-e^{i\theta})^2} \int_{a-i\infty}^{a+i\infty} ds  
\frac{d\ln \zeta(s)}{ds} \int_{-\infty}^{\infty}d(\ln x) e^{(-s_0 - \frac{i}{2}\cot\frac{\theta}{2}+s)\ln x} \nonumber \\
&=& \frac{1}{2\pi i}\int_{-\pi}^{\pi} \frac{d\theta}{2\pi} \frac{e^{-i(m-1)\theta}}{(1-e^{i\theta})^2} \int_{a-i\infty}^{a+i\infty} ds\;  
\delta\left(s-s_0-\frac{i}{2}\cot\frac{\theta}{2}\right)\frac{d\ln \zeta(s)}{ds} \nonumber \\
&=& \int_{-\pi}^{\pi} \frac{d\theta}{2\pi} \frac{e^{-i(m-1)\theta}}{(1-e^{i\theta})^2} 
\left[\frac{d\ln \zeta(s)}{ds}\right]_{s=s_0+\frac{i}{2}\cot\frac{\theta}{2}} \nonumber\\
&=& - \oint \frac{dz}{2\pi i} \,\frac{1}{z^m (1-z)^2}\,\left[\frac{d\ln \zeta(s)}{ds}\right]_{s=s_0 + \frac{1+z}{2(1-z)}}
\label{renBetaShifted} 
\end{eqnarray}
If we take of $s_0=\frac{1}{2}$, so that
\begin{equation*}
s = s_0 + \frac{1+z}{2(1-z)} = \frac{1}{1-z}\qquad\text{and}\qquad \frac{ds}{dz} = \frac{1}{(1-z)^2}
\end{equation*}
we find
\begin{equation}
\beta_m^{\mathrm{ren}} = - \frac{1}{2\pi i}\oint \frac{dz}{z^{m+1}}\,\ln \zeta\left(\frac{1}{1-z}\right)
\label{betamanswer} 
\end{equation}
This should be compared with the coefficients in \cite{Dutta2016byx}
\begin{equation}
\beta_m^{\mathrm{sym}} = - \frac{1}{2\ln 2} \frac{1}{2\pi i} \oint \frac{dz}{z^{m+1}}\,\ln \xi\left(\frac{1}{1-z}\right)
\label{betamPDSD}
\end{equation}
Apart from a constant factor the only difference of this with the renormalized coefficients \cref{renBetaShifted} is that
the symmetric zeta function appear instead of the Riemann zeta function, however, this discrepancy is due to not
incorporating the contribution of the matrix model of $\zeta_{\mathbb{R}}$ discussed in \cref{ssec:MMGamma}.

\section{Normalization of the Resolvent}\label{app:NormResolvent}
The resolvent \cref{pZetaRZed} may be written as 
\begin{align}
R_<^{(p)}(z) &= 1 - \frac{z}{2\ln p}\frac{d}{dz}\ln\left( \sin\left( \ln p\,\frac{1+z}{2i(1-z)}\right)\right) \nonumber\\
&= 1 - \frac{z}{2\ln p} \sum_{n=-\infty}^{\infty} \left( \frac{1}{z-\frac{2n\pi i - \ln p}{2n\pi i + \ln p}} - \frac{1}{z-1}\right)  
\label{ResPoles}
\end{align}
where we have used the infinite product representation of $\sin z = z\displaystyle{\prod_{n=1}^{\infty}} \left(1 - \frac{z^2}{n^2\pi^2}
\right)$. This leads to the density and potential in \cref{pRhoV}.
It can be seen that the poles of \cref{ResPoles} occur pairwise, on the boundary of the unit circle at $z = 1$ and at $z = 
e^{i\theta_n} = \frac{2n\pi i-\ln p}{2n \pi i+\ln p}$. We want to integrate the above in $0 \le \theta\le 2\pi$, along a contour which 
approaches the unit circle from inside (i.e., $|z|\to 1$) and runs counterclockwise. In this limit the contour approaches the singularities 
of the integrand, therefore, the principal value prescription of the integral results in pairs of delta functions with coefficients of
equal magnitude but opposite sign. In this sense, the delta-function peaks in density at non-zero values of $\theta$ cancel the 
negative infinite divergence at $\theta=0$. (This is reminiscent of Coulomb gas with a background charge.) Therefore,
\begin{equation*}
\frac{1}{2\pi i}\oint\frac{dz}{z}R_<^{(p)}(z) = 1
\end{equation*}
i.e., resolvent is normalised correctly. This is consistent with the fact that the resolvent \cref{pZetaRZed} is analytic in the interior of 
the unit disc $|z| < 1$.

If one wants to integrate the density over a finite arc $-\Theta \le \theta \le \Theta$, the right way to interpret the (divergent) 
expression would be to choose a cut-off $N(\Theta)$ so that the sum in \cref{ResPoles} (or in \cref{pRhoV}) is in the range 
$-N(\Theta) \le n \le N(\Theta)$. This way the divergent terms continue to cancel in pairs. With this interpretation the eigenvalue 
density is always positive. Such a cut-off is always possible and independent of the details of the distribution of the singularities. 
Since the poles of the local zeta functions appear as complex conjugate pairs on the line $\mathrm{Re }(s) = 0$, and this symmetry continues to hold after mapping them to the unit circle on the complex $z$-plane, the symmetric choice for the upper and  lower 
cut-off is natural. However, it is clear from \cref{ResPoles} that even with $-N_l(-\Theta) \le n \le N_u(\Theta)$, the divergent terms 
cancel pairwise and results in a positive density. Moreover, in our approach we needed to explicitly construct UMM for only the local 
zeta factors, for which the singularities of the resolvent are known. This construction, therefore, does not depend on the validity of 
the Riemann hypothesis.



\bibliographystyle{hieeetr}
\bibliography{pzeta}{}

\end{document}